\documentclass[twocolumn,times]{aastex631}

\usepackage[T1]{fontenc}
\usepackage{pslatex}
\usepackage{contour}
\usepackage{color}

\usepackage{graphicx}
\usepackage{epsfig} 
\usepackage{natbib}
\usepackage{times}
\usepackage{amssymb,amsmath}
\usepackage{color}
\usepackage{afterpage}
\usepackage{comment}

\definecolor{dark-red}{rgb}{0.8,0,0}
\definecolor{dark-green}{rgb}{0,0.4,0}
\definecolor{dark-blue}{rgb}{0,0,0.8}
\definecolor{dark-magenta}{rgb}{0.8,0,0.8}
\definecolor{orange}{rgb}{1.0,0.6,0}
\definecolor{grey}{rgb}{0.6,0.6,0.6}
\newcommand{\stth}[1]{}
\DeclareMathOperator{\slog}{s-log}

\newcommand{\p}[1]{{{#1}}}
 
\shortauthors{Lionello et al.}
\shorttitle{Global MHD Simulations of the Time-Dependent Corona}

\begin{document}

\title{Global MHD Simulations of the Time-Dependent Corona}

\email{lionel@predsci.com}

\author[0000-0001-9231-045X]{Roberto~Lionello}

\author[0000-0003-1759-4354]{Cooper~Downs}

\author[0000-0002-8767-7182]{Emily~I.~Mason}

\author[0000-0003-1662-3328]{Jon~A.~Linker}

\author[0000-0002-2633-4290]{Ronald~M.~Caplan}

\author[0000-0002-1859-456X]{Pete~Riley}

\author[0000-0001-7053-4081]{Viacheslav~S.~Titov}
\affiliation{Predictive Science Inc., 9990 Mesa Rim Rd., Ste. 170, San Diego, CA 92121, USA} 

\author[0000-0002-6338-0691]{Marc~L.~DeRosa}
\affiliation{Lockheed Martin Solar and Astrophysics Laboratory, 3251 Hanover Street B/203, Palo Alto, CA 94304, USA}

\begin{abstract}
We describe, test, and apply a technique to
incorporate full-sun, surface flux evolution into
%
%incorporate a series of
% photospheric
% magnetic flux evolution maps into a time-dependent, 
an MHD model of the global solar corona. Requiring only maps of the evolving surface flux, our method is similar to that of \citet{2013ApJ...777...76L},
but we introduce two ways to
 correct the electric field at the lower boundary to mitigate spurious
currents. We verify the accuracy of our procedures by comparing 
to a reference simulation, driven with known flows and
electric fields. We then present a thermodynamic MHD calculation lasting one
solar rotation driven by maps from the magnetic flux evolution model of \citet{2003SoPh..212..165S}. The dynamic, time-dependent nature of the model corona is illustrated by examining the evolution of the open flux boundaries and forward modeled EUV emission, which evolve in response to surface flows and the emergence and cancellation flux. Although our main goal is to present the method, we briefly investigate the relevance of this evolution to properties of the slow solar wind, examining the mapping of dipped field lines to the topological signatures of the ``S-Web'' and comparing charge state ratios computed in the time-dependently driven run to a steady state equivalent. Interestingly, we find that driving on its own does not significantly improve the charge states ratios, at least in this modest resolution run that injects minimal helicity. Still, many aspects of the time-dependently driven model cannot be captured with traditional steady-state methods, and such a technique may be particularly relevant for the next generation of solar wind and CME models.

\end{abstract}


%%%%%%%%%%%%%%%%%%%%%%
%%%%%%%%%%%%%%%%%%%%%%
\section{Introduction}
\label{s:intro}
%%%%%%%%%%%%%%%%%%%%%%
%%%%%%%%%%%%%%%%%%%%%%
%
%\cd{I completely redid the intro. Here is are the themes I tried to cover after the 1st paragraph. It could be tighter, sure but mainly let me know if I missed something important.
%\begin{itemize}
%\item Perhaps the most common method for creating full-sun coronal boundary conditions from observations is to create `Synoptic' maps from full-disk magnetograms
%\item The traditional paradigm for modeling the global corona at a given point in time has been  to compute coronal models based on static BCs. examples include PFSS/ADAPT WSA and MHD.
%\item One way to produce models at a higher time cadence is through SFT model output
%\item You can run a bunch of PFSSs but this is missing something, driving preserves the dynamical "memory" from state to state, and allows for the build up of energy as a function of time.
%\item Efield driving is hard and has not yet been explored in full MHD (but the yeats/mackay/weinzerli magnetofrictional stuff does need to be referenced).
%\item To add (optional): The reasoning/justification of the scope of this paper, which is basically the demo of a method with followup papers to come.
%\end{itemize}
%}

Surface magnetic fields are the key observational input to coronal and solar wind models. Used as the inner boundary condition, it is the distribution, polarity, and strength of the radial magnetic field, $B_r$, at the surface that defines the ground state of the corona (the potential field) and largely determines the topology and geometric features of the corona's magnetic field at global scales (coronal holes and streamers). It is also the evolution of the surface flux, which includes the shearing, emergence and cancellation of magnetic features that determines how energy is stored and released in the corona. As such, ground- and space-based observatories routinely measure photospheric and chromospheric magnetic observables, which are then used to create coronal models.

With the exception of the PHI instrument onboard the Solar Orbiter spacecraft \citep{solanki20}, current earth and space-based observatories provide only an Earth-centered view of the photospheric magnetic flux on the Sun. The average rotation period of the Sun as viewed from Earth, known as a Carrington Rotation (CR), is $\approx 27.3$ days, so a traditional approach for constructing model boundary conditions has been to use full-sun Carrington Synoptic maps. These maps are built up by combining the data near central meridian from from full-disk magnetograms as the sun rotates. As such, changes in the average shape and structure of the corona in time have often been characterized in a variety of models on CR-to-CR basis, ranging from simple Potential Field Source-Surface (PFSS) extrapolations to full Magnetohydrodynamic (MHD) calculations. In this modeling paradigm, successive monthly `snapshots' of the corona are computed separately, with each calculation for a given CR being fully independent of the previous case \citep[e.g.,][]{riley06_pfss,luhmann22}.

On the other hand, we know full-well that many of the most intriguing and important phenomena in the corona (e.g., solar flares, jet eruptions, coronal mass ejections, etc.) are inherently driven by surface flux evolution at timescales from minutes to days depending on the spatial scale of interest \citep[][and references therein]{Webb2012,Raouafi2016a,Benz2017}. Similarly, at the largest temporal and spatial scales (i.e. global scales) surface flux evolution is responsible for the formation and evolution of helmet streamers and pseudostreamers over days to months, which is thought to play an important role in the processing of solar wind and ejecta as smaller structures erupt or migrate across their boundaries \citep{Higginson2017,Scott2021,Wyper2021}. How these large-scale closed structures evolve in response to surface changes and subsequently interchange reconnect with nearby open fields has important implications for our understanding of switchbacks measured by Parker Solar Probe \citep{fisk20,zank20,telloni22}, as well as the formation of composition and charge-state variations in the solar wind \citep{zurbuchen02_grl,kepko16}.

%In short, surface flux evolution imprints a dynamical memory in the system as it evolves from state to state, and allows for the build up of magnetic stresses and energy as a function of time. 

In the context of global coronal models, one way to incorporate time-evolution at the inner boundary is by leveraging the outputs of surface flux transport (SFT) models. SFT models describe the time-evolution of magnetic flux over the full-sphere by incorporating various evolutionary processes; typically differential rotation, meridional flow, supergranular diffusion, and random flux emergence. Such models can also assimilate magnetograms from available observatories to produce a continuous approximation of the state of the photospheric magnetic field, also known as a sequence of synchronic maps. Using this approach SFT models have been successful in predicting the evolution of photospheric magnetic fields \citep{1991ApJ...383..431W,2000SoPh..195..247W,2003SoPh..212..165S,2010AIPC.1216..343A,2014ApJ...780....5U}. 

The magnetic flux information from an SFT model can subsequently be processed to create the full-sun boundary condition of $B_r$ for the global coronal magnetic field model. With a time-sequence of synchronic maps, it then becomes possible to model successive states of the corona and heliosphere at a much smaller time-interval than one Carrington rotation. This can be done by running successive (but independent) 3D calculations at the cadence of the synchronic maps \citep[e.g.][]{odstrcil20}, or by driving the model at the inner boundary using electric fields derived from the evolving sequence of $B_r$ \citep[e.g.][]{weinzierl16}. The latter, `driven' approach is physically more attractive as it allows one to capture how surface flux evolution surface flux evolution imprints a dynamical `memory' in the system as it evolves from state to state, and allows for the build up of magnetic stresses and energy in time at the correct dynamical timescales.

%the dynamical timescales for field evolution, which includes both energy storage and release as well as topological change. 

On the other hand, deriving an appropriate driving electric field from a sequence of $B_r$ maps is both challenging and not fully constrained \citep[][]{fisher10,cheung12,yeates17,lumme17}. For global coronal coronal field models, such driving was studied with the simplified magnetofrictional approach by \citet{weinzierl16} who focused on the importance of the non-inductive freedom in the boundary electric field. While convenient for studying energy injection and storage, magnetofrictional models fundamentally cannot capture the dynamical timescales of the system (set by the magnetoacoustic speeds), non-force free processes (such as eruptions), nor how the 3D plasma state of the corona and solar wind evolve in tandem with surface flux changes (flows, density, temperature). Another challenge for full-sun, global driving is the fact that data assimilation in SFT models, which ingests new measurements from the Earth-Sun line as the Sun rotates, will instantaneously overwrite existing flux.
These imposed changes imprint both an unphysical dynamical timescale and forcing on the system as well as a floating magnetic monopole that must (typically) be corrected by some means.

In this light, we describe our efforts to develop and test a suitable boundary driving approach for global MHD models of the solar corona. In \S\ref{sec-evolution}, using a technique similar to \citet{2013ApJ...777...76L}, we illustrate how a full electric field may be expressly determined from time-evolving $B_r$ maps and the flow profile from an SFT model. 
At the boundary, the electric field obtained through our formulation evolves the magnetic field smoothly at the code time-step, thus avoiding the instantaneous overwriting of the flux. We also discuss additional corrections that can help eliminate spurious currents at the inner boundary of the MHD model. In \S\ref{sec-test} we test and compare the various approaches on an idealized case where the true surface flows are known.

Next in \S\ref{sec-demonstrate} we demonstrate the approach in a full thermodynamic MHD calculation for one-month time of coronal evolution, which is driven by a sequence of magnetic flux maps provided by the Lockheed Evolving Surface-Flux Assimilation Model \citep[ESFAM,][]{2003SoPh..212..165S}. This particular ESFAM run does not assimilate data, but is instead designed to yield Sun-like evolution with an average strength sunspot cycle at solar minimum conditions. This provides a smoothly evolving full-sun SFT dataset without the typical artifacts of data assimilation. To characterize the results, we calculate EUV synthetic emission images, as well as maps of coronal holes locations, the squashing factor \citep[$Q$,][]{2007ApJ...660..863T}, dips in magnetic field lines, and fractional charge states.  To evaluate the importance of time-dependent evolution, we also compare these results with those of steady-state models. Finally, we conclude in \S\ref{sec-dis} by discussing the our results in the context of the solar wind as well as the relevance of these techniques for future applications.

%%%%%%%%%%%%%%%%%%%%%%
%%%%%%%%%%%%%%%%%%%%%%
\section{PARAMETERIZING FULL-SUN BOUNDARY EVOLUTION IN MHD}
\label{sec-evolution}
\subsection{The MAS MHD Model}

To develop, test, and study methods for evolving the magnetic flux at the inner boundary of an MHD model, we employ the  
Magnetohydrodynamic Algorithm outside a Sphere (MAS) code. MAS is designed to model the 
global solar atmosphere from the top of the chromosphere to Earth and beyond
 and  has been used extensively to study coronal structure 
\citep{1999PhPl....6.2217M,1999JGR...104.9809L,2009ApJ...690..902L,
2013Sci...340.1196D,2018NatAs...2..913M}, 
coronal dynamics \citep{2005ApJ...625..463L,2006ApJ...642L..69L,
2011ApJ...731..110L} and CMEs \citep{2003PhPl...10.1971L,2013ApJ...777...76L,
2018ApJ...856...75T}. 
MAS solves the resistive, thermodynamic MHD equations in spherical 
coordinates $(r,\theta,\phi)$ on structured nonuniform meshes. 
Magnetosonic waves are treated semi-implicitly, allowing us to use large 
time steps for the efficient computation of long-time evolution. {The semi-implicit method is not harmful for obtaining time-dependent solutions, as it
introduces dispersive effects only for processes occurring on time-scales shorter than or equal to the time step \citep{1987JCoPh..70..330S}. Given that the flows that drive evolution of flux at the surface are much smaller than the typical coronal flow speeds that set the CFL limit and  time step in the model, the semi-implicit method is more than suitable for the time-dependently driven calculations described here.}

The present version of MAS allows for several modes of operation that govern which terms are solved for and/or added to the MHD equations. Here we use the term `thermodynamic MHD` to indicate that MAS solves for additional transport terms that describe energy and momentum flow in the solar corona and solar wind \citep[coronal heating, parallel thermal conduction, radiative 
loss, and Alfv\'e{}n wave acceleration; as fully described in  
Appendix A of][]{2018ApJ...856...75T}.

The latest version of the thermodynamic mode in MAS also includes a physics-based specification of
the coronal heating term through a Wave-Turbulence-Driven (WTD) phenomenology.
In this approach, additional equations are solved to capture the macroscopic
propagation, reflection, and dissipation of low-frequency Alfv\'enic
turbulence. The physical motivation underlying the WTD approach is that
outward and reflecting Alfv\'e{}n waves interact with one another, resulting in
their dissipation and heating of the corona
\citep[e.g.,][]{1996JGR...10117093Z,2007ApJ...662..669V}. This follows related
works, where the general formalism for the propagation of Alfv\'en waves
\citep[e.g.,][]{1980JGR....85.1311H,1996JGR...10117093Z,2012ApJ...756...21Z} is
usually approximated to produce tractable equations for the propagation of the
energy density or the amplitude of the Alfv\'en waves
\citep[e.g.,][]{1993A&A...270..304V,1999ApJ...523L..93M,2001ApJ...548..482D,
2007ApJS..171..520C, 2007ApJ...662..669V, 2008JGRA..113.8105B,
2009ApJ...707.1659C, 2011ApJ...727...84U, 2012ApJ...745....6J,
2013ApJ...764...23S, 2014ApJ...782...81V, 2015ApJ...806...55O}.
A full description of
 the MAS-WTD model and equations solved is provided in the supplementary 
materials of \citet{2018NatAs...2..913M}.
 
We have also recently incorporated into
MAS a non-equilibrium ionization module to advance the
 fractional charge states of minor ions according to the model
of \citet{2015A&C....12....1S}:
\begin{equation}
\begin{split}
\frac{\partial {}_{Z}F^i}{\partial t} + \mathbf{v} \cdot \nabla {}_{Z}F^i 
=
n_e\left [ {}_{Z}C^{i-1} {}_{Z}F^{i-1} \right . \\
\left . - \left ( {}_{Z}C^i +{}_{Z}R^{i-1} \right
 ) {}_{Z}F^i + {}_{Z}R^{i}  {}_{Z}F^{i+1} \right ]. \label{e:cs}
\end{split}
\end{equation}
For an element with atomic number $Z$, ${}_{Z}F^i(r,\theta,\phi)$ indicates the
fraction of ion ${i+}$ ($i=0,Z$) with respect to the total at a grid point:
\begin{equation}
\sum_{i=0}^{Z} {}_{Z}F^i = 1.
\end{equation}
For each element, the ion fractions are coupled through the ionization,
 ${}_{Z}C^{i}(T)$, and recombination, ${}_{Z}R^{i}(T)$, rate coefficients
derived
from the CHIANTI atomic database
\citep{1997A&AS..125..149D,2013ApJ...763...86L}. Here
$T$, $n_e$, and $\mathbf{v}$ are respectively the temperature, number
density, and the velocity of the plasma.
This module has
been tested in our hydrodynamic 1D, WTD wind code \citep{2019SoPh..294...13L}. {A similar time-dependent 3D model of charges states of minor ions is shown in \citet{2022ApJ...926...35S}.}

\subsection{Incorporating Magnetic Flux Maps}
\label{ss:corrections}
To drive the magnetic field evolution in MAS,
we can evolve the radial component
of the magnetic field at the boundary  using a technique similar to that
described by \citet{2013ApJ...777...76L}.   We specify the
tangential electric
field at the boundary  $\mathbf{E}_{t0}(\theta,\phi,t)$  as
\begin{equation}
\mathbf{E}_{t0}=\nabla_t \times \Psi \mathbf{\hat{r}}+\nabla_t  \Phi . 
\end{equation}
The potential $\Psi $ controls the evolution of the normal component of the 
magnetic field $B_{r0}$,
\begin{equation}
\nabla^2_t \Psi = \frac{\partial B_{r}}{\partial t}. \label{e:psi}
\end{equation}
We use a sequence of full-sun $B_r(\theta,\phi)$ maps {in time} to specify the evolution of $\Psi$ using
Eq.~(\ref{e:psi}). {Since the temporal cadence of the input maps will generally be much slower than the MHD model time step, we must interpolate $B_r$ in time at every step. Because linear interpolation of the $B_r$ maps in time implies a discontinuous $\partial B_{r}/ \partial t$ as we shift from the interval between one pair of maps to the next, one must use a higher order interpolation scheme to ensures a smooth step-to-step evolution of the electric field at the inner boundary. Here we use a simple piece-wise cubic Hermite interpolation scheme \citep{1980SJNA...17..238F}, to ensure continuity in at least the first derivative and preserve monotonicity in the evolution of $B_r(t)$ at all points within in the time-interval between maps.}

{Next}, the potential $\Phi(\theta,\phi,t)$ can be completely 
specified if the flows that led to
the evolution are known, i.e., we can find $\Phi $ from the equation:
\begin{equation}
\nabla^2_t \Phi=-\nabla_t \cdot(\mathbf{v}
\times \mathbf{B})_t . \label{e:phi}
\end{equation}
In general however, the full-sun distributions of the complete flow and magnetic field vectors responsible for map-to-map $B_r$ changes are not typically available from observations or SFT models that assimilate observational data. Instead we can incorporate the known large-scale flows (differential rotation and meridional flows) to at least include their contribution to the transverse electric field. In other words, while the $\Phi$ potential solve ensures that we can evolve flux to exactly match the $B_r$ component of the maps, the magnetic fields that evolve and emerge will generally have less shear and twist than is observed (especially for ARs). On the other hand, this method has a significant advantage over more complex or localized techniques in that vector magnetic field observables and/or flow correlation tracking are not required. Furthermore, additional energization can easily be explored through the freedom in the $\Phi$ potential\footnote{See the MAS energization examples in \citet{yeates18_ssrv,2018NatAs...2..913M} where a full-sun $\Phi$ potential is specified in a semi-automated procedure to emerge shear along arbitrary neutral lines.}, which will be explored in future work.

At this point the $\Psi$ and $\Phi$ potentials only provide information about the transverse electric field. Our next aim is to find a full boundary electric field, $\mathbf{E}_{0}$, that
is (1) minimally diffusive and causes (2) minimal boundary layers. 
A possible solution is to prescribe the condition 
 of ideal MHD at the lower boundary (method \textsf{A}):
\begin{equation}
\mathbf{E}_{0} \cdot \mathbf{B}_{0}= E_r B_r + \mathbf{E}_{t0} \cdot \mathbf{B}_{t0} = 0,
\end{equation}
where  $t$ means the 
component tangential to the boundary 
 of the $\mathbf{B}_{0}$ and $\mathbf{E}_{0}$  fields.
 From that it follows that
\begin{equation}
E_r = - \frac{\mathbf{E}_{t0} \cdot \mathbf{B}_{t0}}{B_r},
\end{equation}
which can be regularized using  a small parameter $\epsilon$ as
\begin{equation}
E_r^* = - \frac{(\mathbf{E}_{t0} \cdot \mathbf{B}_{t0}) B_r}{B_r^2 +
 \epsilon^2}.
\label{eq-er}
\end{equation}
This method has the disadvantage of introducing artificial reconnection
at the polarity inversion line (PIL), where $B_r=0$. 

Building upon method \textsf{A}, 
we can compute an additional correction to the boundary
electric field, $\tilde{\mathbf{E}}_{0}$, such that
  $ \tilde{\mathbf{E}}_{0} \cdot \mathbf{B}_0 \simeq 0 $ everywhere by slightly modifying $\mathbf{E}_{t0}$ and $E_r^*$ near the neutral line 
(method \textsf{B}).
To define $\tilde{\mathbf{E}}_{0}$,
 we use 
\begin{equation}
 \mathbf{v}_\perp^* =  \frac{\mathbf{E}_{0} \times \mathbf{B}_{0}}{B^2},
\end{equation}
which is the flow associated with $\mathbf{E}_{0}$ as  calculated 
with \textsf{A}:
\begin{equation}
\tilde{\mathbf{E}}_{0}= - \mathbf{v}_\perp^* \times \mathbf{B}_{0}
\end{equation}
We obtain
\begin{eqnarray}
\tilde{E}_{r_0} &=& \left ( 1 + \frac{\epsilon^2}{B^2} \right )
 \frac{\mathbf{E}_{t_0}\cdot \mathbf{B}_{t_0}}{B^2_{r_0}+\epsilon^2} B_{r_0}, \\
\tilde{\mathbf{E}}_{t_0} &=& \mathbf{E}_{t_0} - \frac{ \epsilon^2 
\mathbf{E}_{t_0}\cdot \mathbf{B}_{t_0}}{B^2_0 \left ( B^2_{r_0} + \epsilon^2 
\right )} \mathbf{B}_{t_0}.
\end{eqnarray}

Whether method \textsf{A} or \textsf{B} is used, 
we derive again the boundary flow, $ \tilde{\mathbf{v}}_0$, using the
boundary electric field 
expression provided by  either  method,
\begin{equation}
\tilde{\mathbf{v}}_0 = \frac{  \tilde{\mathbf{E}}_{0}  \times
\mathbf{B}_{0} }{ \epsilon^2 +B^2_0},
\label{eq-v0}
\end{equation}
where we add a  small parameter $\epsilon$ to regularize the flow at the neutral line.

Only for method \textsf{B}, we also add a small resistivity value  at the boundary  PIL:
\begin{equation}
\tilde{\eta}  \propto \left | 
 \frac{ \partial \sin \theta \tilde{E}_{\phi}}{\partial \theta}
-\frac{1}{\sin \theta}\frac{ \partial \tilde{E}_{\theta}  }{\partial \phi }
-  \left (
 \frac{ \partial \sin \theta E_{\phi}}{\partial \theta}
-\frac{1}{\sin \theta}\frac{ \partial E_{\theta}  }{\partial \phi }
\right )
\right |,
\label{eq-method-b-resistivity}
\end{equation}
where we dropped the subscript 0 from the notation.

\subsection{Asymptotic Values at the PIL}
Because each method requires some form of regularization near the PIL, it follow that we should seek a robust dimensionless formulation that is equally appropriate all types of broad/compact and weak/strong flux-distributions on the Sun. 

We can rewrite Eq.~(\ref{eq-er}) in terms of 
$f=\epsilon^2/B_r^2 $  and obtain

\begin{equation}
\frac{|E_r^0|}{|\mathbf{E}_{t0}|} \propto \frac{|\mathbf{B}_{t0}|} {|B_r|(1+f)}.
\label{eq:trunc_rewrite}
\end{equation}

The right-hand side (RHS) of Eq.~\ref{eq:trunc_rewrite} can be easily computed 
at any state of the simulation, regardless of $dB_r/dt$ or the
 full electric field itself. We seek a definition of $f$ such that $E_r$ is 
bounded 
near the PIL 
and $f$ is controlled by a dimensionless constant, unlike $\epsilon$, which 
 is in units of $B$. Let us therefore find a form of $f$ so that the RHS
of Eq.~\ref{eq:trunc_rewrite} will reach a fixed, asymptotic value
when $B_0\to0$. We simply define $f$ as
\begin{equation}
f=A\frac{|\mathbf{B}_T|}{|B_r|},
\label{eq:f_newdef}
\end{equation}
where $A$ is a dimensionless constant, and $1/A$ sets the largest value
for the LHS of Eq.~\ref{eq:trunc_rewrite}. The new form assumed by 
Eq.~(\ref{eq-er}) is thus the following:
\begin{equation}
E_r^* = -\frac{\mathbf{E}_{t0}\cdot\mathbf{B}_{t0}\ \text{sign}(B_r)}{|B_r|+A|\mathbf{B}_{t0}|}.
\label{eq:trunc_mas}
\end{equation}

Similarly, we seek a new formulation for the boundary flow
 $\tilde{\mathbf{v}}_0$ in
 Eq.~(\ref{eq-v0}) 
that will converge to a  meaningful asymptotic value for
$\tilde{\mathbf{v}}_0$,  when $B_0 \to 0 $  at the PIL. We rewrite Eq.~(\ref{eq-v0})  as 
\begin{equation}
\tilde{\mathbf{v}}_0 =
 \frac{\tilde{\mathbf{E}}_0\times\mathbf{B}_0}{B_0^2(1 + g)},
\label{eq:trunc_v_new}
\end{equation}
where $g$ is a dimensionless quantity. We proceed heuristically by comparing
$\tilde{\mathbf{v}}_0$ to the local
 sound speed, calculating the magnitude, and dividing by $c_s$ to obtain
\begin{equation}
\frac{|\tilde{\mathbf{v}}_0|}{c_s} = \frac{|\tilde{\mathbf{E}}_0||\mathbf{B}_0|\sin \theta}{c_sB^2_0(1 + g)}=\frac{\tilde{E}_0\sin \theta}{c_s B_0 (1 + g)}.%\simeq \frac{1}{D}
\label{eq:trunc_v_new_ratio}
\end{equation}
Then we pose \begin{equation}
g=D\frac{\tilde{E}_0}{c_s B_0},
\label{eq:gfunc}
\end{equation}
where $D$ is a dimensionless constant, and insert this into Eq.~(\ref{eq:trunc_v_new_ratio}). The ratio simplifies as
\begin{equation}
\frac{\tilde{\mathbf{v}}_0}{c_s} = \frac{\tilde{E}_0\sin \theta}{c_s B_0 + D \tilde{E}_0},
\label{eq:vperp_heuristic}
\end{equation}
which converges to $1/D$ as $B_0\to 0$. In other words,
  $1/D$ specifies the maximum fraction of $|\tilde{\mathbf{v}}_0|$ relative to the local sound
speed at the boundary.
The new formulation of $\tilde{\mathbf{v}}_0$ thus becomes
\begin{equation}
\tilde{\mathbf{v}}_0
 = \frac{\tilde{\mathbf{E}}_0\times\mathbf{B}_0}{B^2_0 + {D \tilde{E}_0 B_0}/{c_s}}.
\label{eq:trunc_v_new_mas}
\end{equation}

%%%%%%%%%%%%%%%%%%%%%%
%%%%%%%%%%%%%%%%%%%%%%
\section{TEST OF THE CORRECTIONS}
\label{sec-test}
To evaluate the performance of the correction methods in
\S\ref{ss:corrections}, we can apply them to a solution where the true electric
field that evolved the boundary is known. To this end, we apply them to time-dependent simulations driven
with the magnetic flux maps extracted from the 3D MHD coronal solution
of \citet{2005ApJ...625..463L,2020ApJ...891...14L}.
These employed the polytropic model of \citet{1999JGR...104.9809L} to study
the effects of 
 differential rotation on a
photospheric magnetic field distribution similar to that of
 \citet{1996Sci...271..464W}. This consists of a global dipole distribution
with a bipolar active region superimposed (see also Fig.~\ref{fig-2d-testruns}).
After a relaxation period lasting till $t=150 $ code units (1 CU = 1445.87 s), 
they turned the following surface flow on:
\begin{equation} 
\omega(\theta) = 13.39 -2.77 \cos^2 \theta \deg \mathrm{day}^{-1},
\label{eq-diff_rot}
\end{equation}
which is 10 times the value of \citet{1996Sci...271..464W}.  We repeated this
run in the same manner, evolving the surface field directly using the prescribed $\mathbf{E}=-\mathbf{v}\times\mathbf{B}$ from the driving flow (only $v_\phi$ in this case), and $\mathbf{B}_0$ at a given timestep. We label this run with \textsf{O} (original), and sample the values of $B_r$ on the solar surface with a 1 CU cadence.

Then we use the surface $B_r$ maps from \textsf{O} to run four simulations using 
Eqs.~(\ref{e:psi}-\ref{e:phi}) to prescribe the electric field at the 
$r=R_\odot$ surface: the first run is
uncorrected (\textsf{U}, i.e.\ neither of the 
 methods of \S \ref{ss:corrections}
is applied but the `true' radial electric field is used, $E^0_r = \omega R_\odot B^0_\theta$), the second run (\textsf{E}) is erroneously driven with $E_r^0=0$,
 the third run is corrected with method \textsf{A}, and
the fourth is corrected with \textsf{B}.

%=======================================================
% Time History plots for the test runs
%=======================================================
\begin{figure*}[t]
\begin{center}
\includegraphics[width=0.8\textwidth]{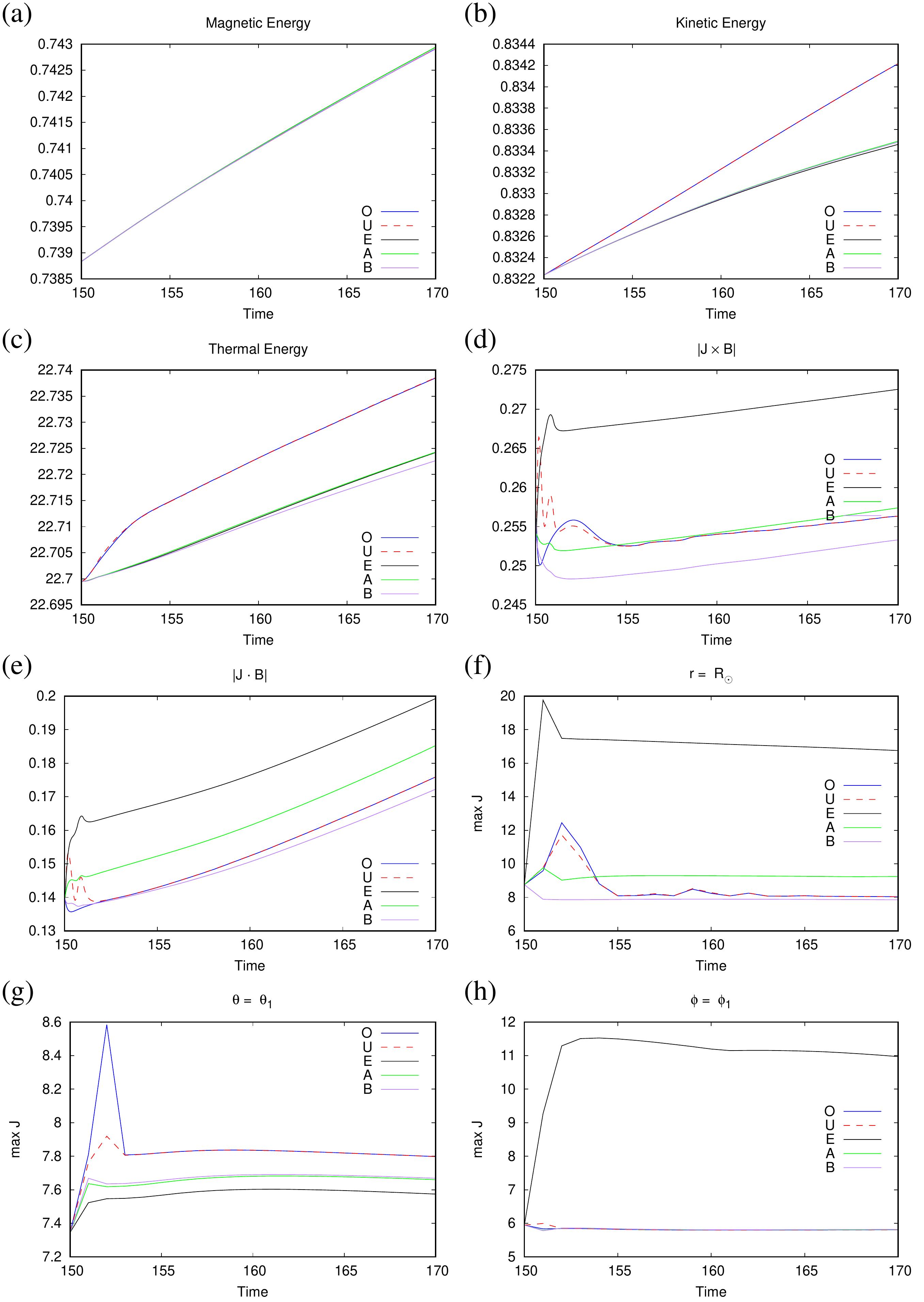}
\end{center}
\caption{Time histories of the simulations respectively labeled
 \textsf{0}, \textsf{U}, \textsf{E}, \textsf{A}, and \textsf{B}  
in \S\ref{sec-test}.
Volume integrals of the magnetic energy (a), of the kinetic energy (b),
 of the thermal energy (c), of $|\mathbf{J} \times \mathbf{B}  |$ (d), 
and of $|\mathbf{J} \cdot \mathbf{B} | $ (e). Maximum current density
$J$ evaluated on  the surfaces $r=R_\odot$ (f), $\theta = \theta_0$ (g),
and $\phi = \phi_0$ (h). The locations of $\theta_0$ and $\phi_0$ are shown in the upper left panel Fig.~\ref{fig-2d-testruns}. 
}
\label{fig-hfile}
\end{figure*}

%=======================================================
% 2D Maps of the test run quantities
%=======================================================
\begin{figure*}[hbtp]
\begin{center}
\includegraphics[width=0.99\textwidth]{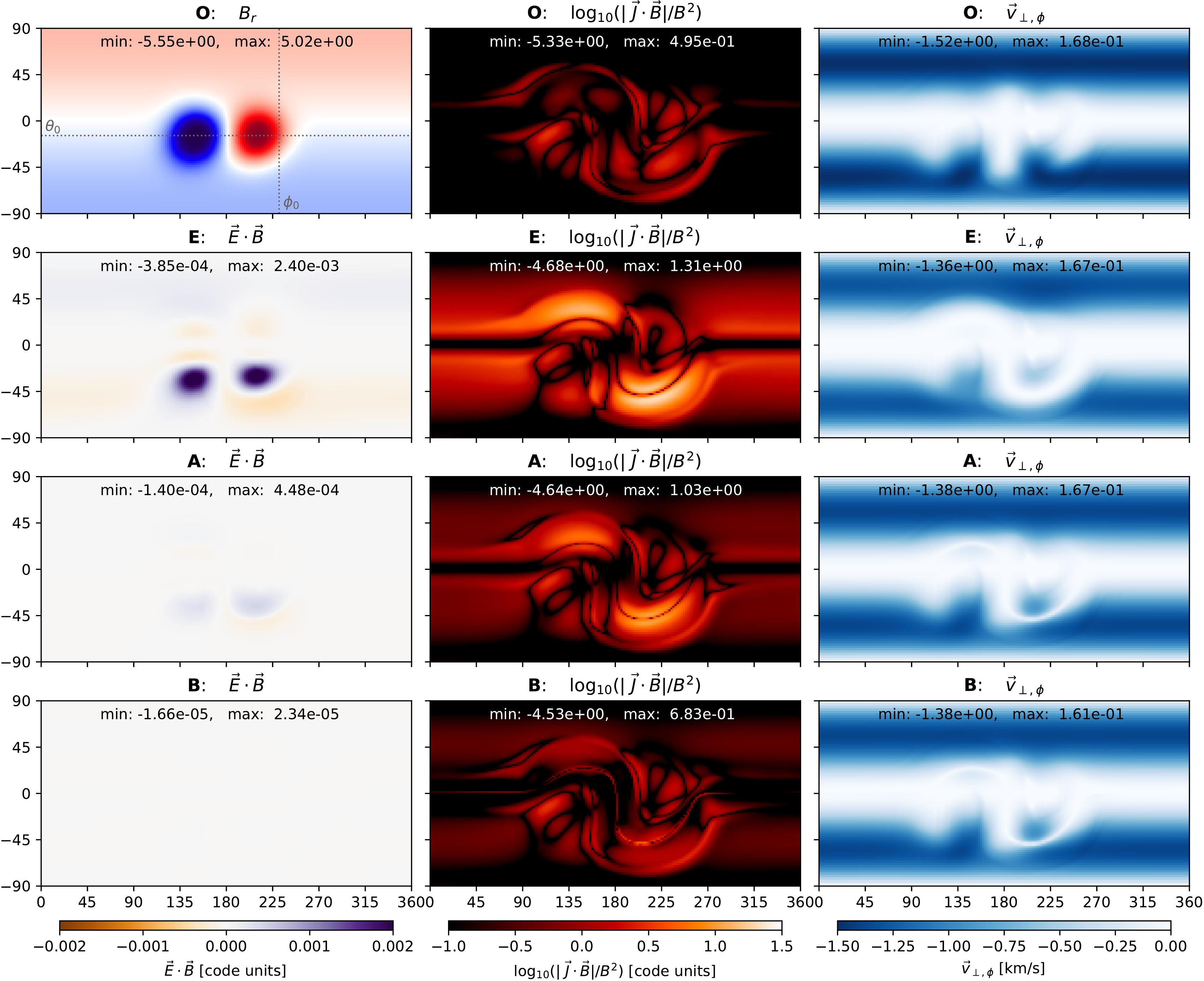}
\end{center}
\caption{2D maps of surface quantities at the final step ($t=170$ CU) for test cases \textsf{O}, \textsf{E}, \textsf{A}, and \textsf{B}. The upper left panel shows $B_r$ at this time for reference and the positions of the $\theta_0$ and $\phi_0$ cut planes used in Fig.~\ref{fig-hfile} are indicated with the dotted gray lines. The remaining left panels show $\mathbf{E}\cdot\mathbf{B}$ for each case, where $\mathbf{E}$ is the driving electric field. The middle column shows $\log_{10}(|\mathbf{J}\cdot\mathbf{B}|/B^2)$ for all four runs. The right column shows the $\phi$ component of the driving velocity that is perpendicular to $\mathbf{B}$ ($v_{\perp,\phi}$). For each map, the  minimum and maximum of the plotted quantity at the surface are shown.}
\label{fig-2d-testruns}
\end{figure*}

%=======================================================
In Fig.~\ref{fig-hfile} we present the time histories of the
four runs from the inception of photospheric flows at $t=150$ until $t=170$. Panels (a-f) show quantities integrated over the whole
computational domain. With the exception of currents, the \textsf{U} run presents values coincident with those
of the  \textsf{O} run in all panels.
 The magnetic energy (Fig.~\ref{fig-hfile}a) of
the four runs are also practically indistinguishable. However, following
 the  introduction of differential rotation at 
$t=150$, the kinetic
energy (Fig.~\ref{fig-hfile}b) of the  \textsf{A} and \textsf{B} driven runs 
 is $\sim 1\%$ smaller than the reference \textsf{O} run.
This behavior is replicated in the time history of the magnetic energy in
Fig.~\ref{fig-hfile}c. The integrated Lorentz force (Fig.~\ref{fig-hfile}d)
shows a jump at  ($t=150$) for all runs, with  the curve of the
  \textsf{O} and \textsf{U}  runs 
bracketed by those of  
of the corrected runs \textsf{A} and \textsf{B}. The last integrated
quantity $|\mathbf{J}\cdot\mathbf{B}|$ measures the deviation from 
 ideal MHD. Similarly to (d), \textsf{U} and \textsf{O} are between 
\textsf{A} and \textsf{B}. The erroneous run, \textsf{E} has larger
values of the integrated Lorentz force and parallel current.

We also examine the history
of the maximum current density $J$ on three orthogonal cut-planes in spherical coordinates:
the $r=R_\odot$ surface encompasses the whole
active region; the $\theta = \theta_0$ surface intersects
the active region; the $\phi = \phi_0$ lies west of the positive polarity.
 When 
the flows are turned on at $t=150$, we notice in panel (f) 
 a sudden spike  on the  $r=R_\odot$
surface  for the \textsf(O) and  \textsf{U} run, while the enhancements for 
\textsf{A} and \textsf{B} are much smaller. As it was the case 
for the global measurements in (d) and (e),
the \textsf(O) and  \textsf{U} values soon become bracketed by those
 of \textsf{A} and \textsf{B}. Run  \textsf{E} shows high values of the
maximum currents on the $r=R_\odot$ and $\phi = \phi_0$ surfaces.
 The (g) and (h) panels  shows that, with the exception
of \textsf{E}, all the driven runs
manage to reproduce the maximum $J$ on $\theta = \theta_0$  and 
$\phi = \phi_0$  surfaces
 in substantial agreement with the 
\textsf{O} run.

To convey a more intuitive sense of how the corrections operate, we show full-sun surface maps of various quantities at the final state of runs \textsf{O}, \textsf{E}, \textsf{A}, and \textsf{B} in Fig.~\ref{fig-2d-testruns} (skipping \textsf{U} because it is essentially the same as \textsf{O}).  First, to illustrate the shape and location of the PIL near the large bipolar region, the upper left panel shows the full-sun $B_r$ distribution. The remaining panels in the left column show $\mathbf{E}\cdot\mathbf{B}$ for the computed electric field for runs \textsf{E}, \textsf{A}, and \textsf{B} ($\mathbf{E}\cdot\mathbf{B}$ is zero for \textsf{O} by construction). As expected, $\mathbf{E}\cdot\mathbf{B}$ is largest when $E_r$ is ignored altogether in run \textsf{E} and the signature is strongest near the largest fields along the the PIL. Method \textsf{A} reduces the maximum $\mathbf{E}\cdot\mathbf{B}$ compared run \textsf{E} by just over a factor of 5, while method \textsf{B} effectively eliminates it as intended.

The middle column of Fig.~\ref{fig-2d-testruns} illustrates how field-aligned currents build up at the boundary in each case, showing a map of $\log_{10}(|\mathbf{J}\cdot\mathbf{B}|/B^2)$. Compared to the reference case \textsf{O}, all runs introduce some level of additional currents but the features change from case to case. As expected, by not applying any $\mathbf{E}\cdot\mathbf{B}=0$ correction, run \textsf{E} builds up the largest boundary layer, especially near the main PIL of the two polarities. Method \textsf{A} attenuates this current signature by a factor of two or more (note the $\log_{10}$ scaling of the min/max values), while method \textsf{B} attenuates this signal even further. However, right at the PIL for run \textsf{B} we see a slight enhancement of $\mathbf{J}\cdot\mathbf{B}$ compared to the other runs, which is the signature of the small resistivity that is active where $B_r\to0$ that is needed to ensure the $B_r$ evolution matches the driving sequence of maps (Eq.~\ref{eq-method-b-resistivity}).

Lastly, the right panels of Fig.~\ref{fig-2d-testruns} show maps of the longitudinal component of the velocity field perpendicular to $\mathbf{B}$ at the inner boundary, $v_{\perp,\phi}$. For run \textsf{O} this is simply the component of the differential rotation flow that actually moves the field eastwards, while for the remaining cases this component is determined from Eq.~\ref{eq:trunc_v_new_mas}. For \textsf{E}, \textsf{A}, and \textsf{B} we see that the true driving flow is largely recovered everywhere except near the PIL of the flux concentrations. Methods \textsf{A} and \textsf{B} do a better job than run \textsf{E} near the PIL, especially in filling in the strong patches of $v_{\perp,\phi}$ to the southeast and northwest of the central PIL segment. That said, on the opposite side of the sun where dipolar flux-distribution is basically symmetric north south (and thus $B_r$ is not changing) the maximum strength of $|v_{\perp,\phi}|$ is not quite recovered. This stems from the fact that the PIL for the global dipole is very broad, such that the truncation function for determining $E_r^0$ (Eq.~\ref{eq:f_newdef}) is not quite zero even at mid-latitudes.

In summary, these tests runs illustrate the inherent limitations of recovering a suitable electric field to drive surface flux evolution from limited information (a sequence of full-sun $B_r$ maps in time) as well as the pros and cons of the various methods. When the full surface flow profile is known, the combination of $\mathbf{E}_t$ derived from the $\Psi$ and $\Phi$ potential solves and the true $E_r$, which can be computed if the exact surface flow is known, reproduces the true solution (runs \textsf{O} vs \textsf{U}). When $E_r$ is not available, which will be generally true when only the full-sun evolution of $B_r$ and the average macroscopic flows are known (i.e. flux-transport models assimilating observations), it must be computed from available information (runs \textsf{A} vs \textsf{B}), or ignored (run \textsf{E}).

Although both correction methods \textsf{A} and \textsf{B} appear to give similar results, method \textsf{B} is able to completely eliminate the non-ideal part of the electric field, which may help it perform better when strong magnetic flux is carefully emerged (e.g., during the simulations of CMEs). On the other hand, the advantages of method \textsf{B} comes at the at the cost of an additional free parameter ($\tilde{\eta}$) and concentrating the resistive boundary layer right at the PIL. In favor of smoothness and simplicity, we thus opt to use method \textsf{A} in our first application of this approach to a more realistic case, which is described in the following section.

%%%%%%%%%%%%%%%%%%%%%%
%%%%%%%%%%%%%%%%%%%%%%
\section{TIME DEPENDENT WTD MODEL}
%=======================================================
\begin{figure*}[t]
\begin{center}
\includegraphics[width=0.8\textwidth]{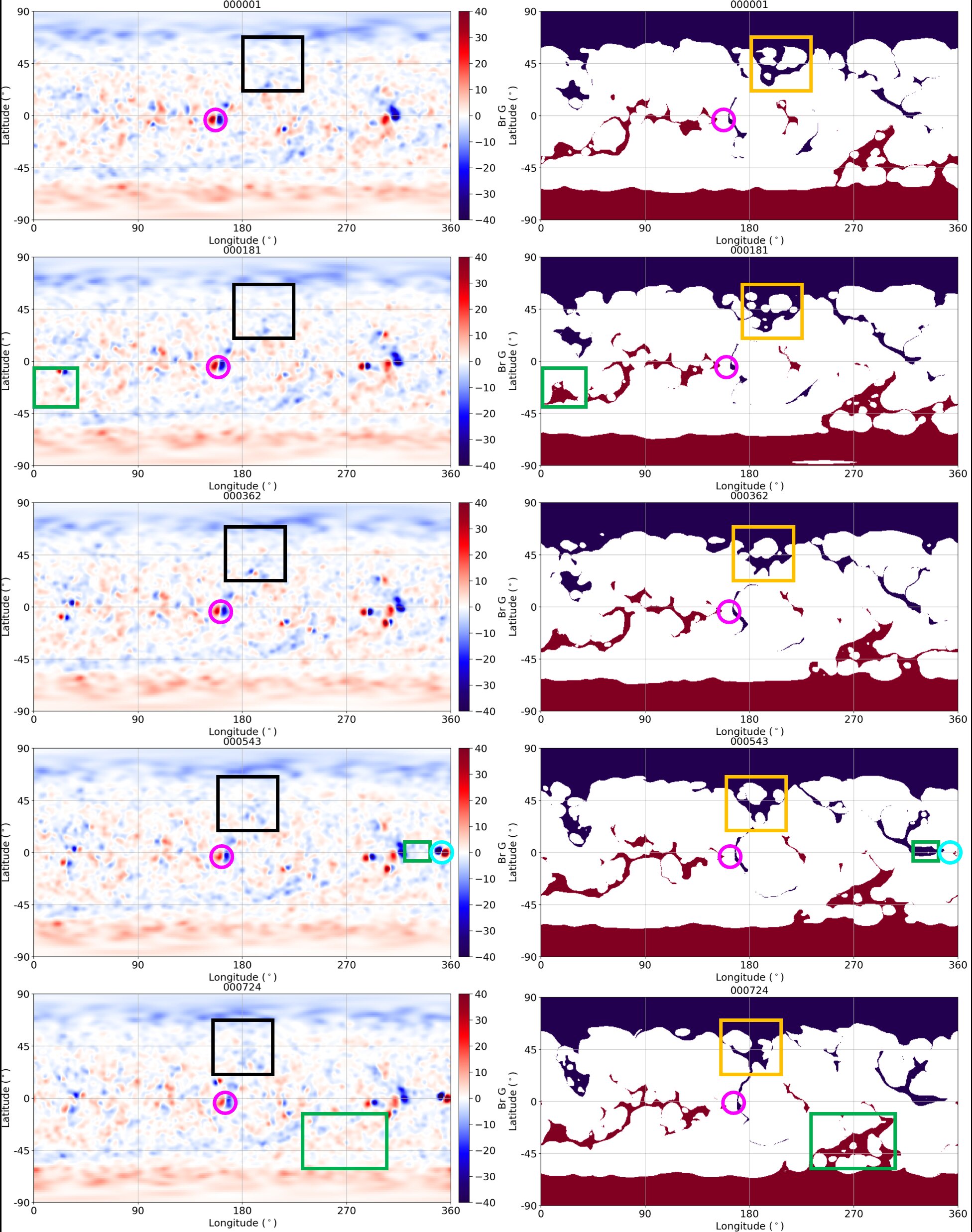}
\end{center}
\caption{
Evolution of the photospheric magnetic field and the open flux boundaries
 over one
month in the time-evolving MHD model. 
The left hand frames show $B_r$, derived from the ESFAM
flux transport model, as a latitude-longitude map. These are the boundary 
maps for the calculation.
The right-hand frames show open/closed boundaries (coronal holes) in the same format, with dark
red indicating outward (positive) polarity, dark blue showing inward (negative) polarity, and white
indicating closed-field regions. Five time instances are shown, from top to bottom: 
$t=0$, $t \simeq 180~\mathrm{hrs}$, $t \simeq 360~\mathrm{hrs}$, $t \simeq 540~\mathrm{hrs}$, and $t \simeq 720~\mathrm{hrs}$. The box on every frame (black on the left, gold on the right for visibility) highlights a persistent open field region, while the magenta circle near the center of each frame indicates a bipole which undergoes decay. Other rectangles and circles denote more transient open field and flux emergence regions, respectively. A 7-second animation of this figure is available online (\href{https://www.predsci.com/corona/tdc/animations/RL_2023_Fig3.mp4}{www.predsci.com/corona/tdc/animations/RL\_2023\_Fig3.mp4}) showing both maps for the duration of the simulation time.}

\label{fig-br_ch}
\end{figure*}

\begin{figure*}[t]
\begin{center}

\includegraphics[width=0.75\textwidth]{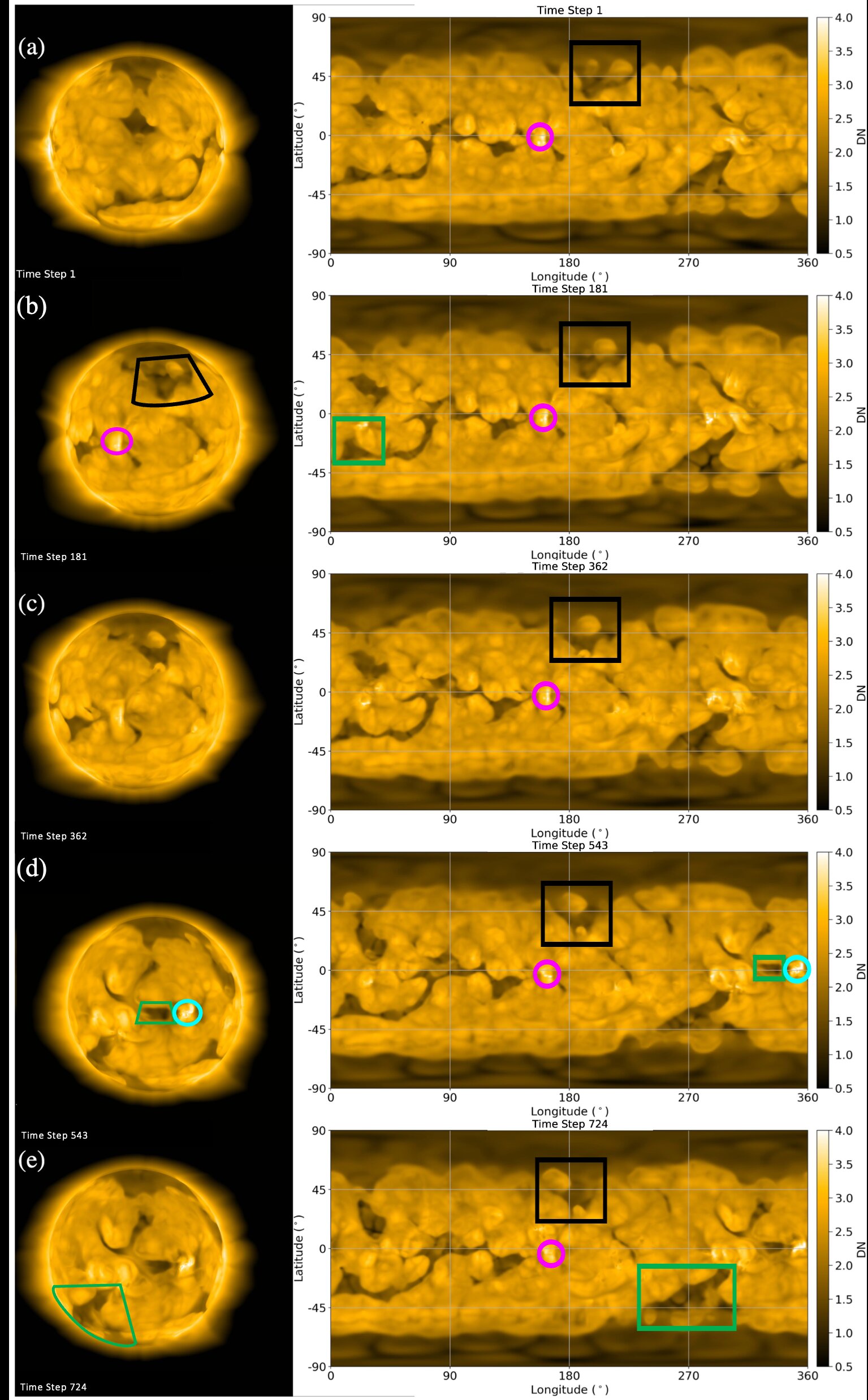}
\end{center}
\caption{
The emission in the AIA 171 \AA\ channel 
 over one
month in the time-evolving MHD model. 
The left hand frames show the Sun from the point of view of
an observer on Earth, as the star completes a full
rotation around its axis.
The right-hand frames show a projection of the emission as 
a latitude-longitude map; the annotations are analogous to those from the previous figure.
Five time instances are shown, corresponding to those 
of Fig.~\ref{fig-br_ch}: (a)
$t=0$ (b) $t \simeq 180~\mathrm{hrs}$
 (c) $t \simeq 360~\mathrm{hrs}$  (d)  $t \simeq 540~\mathrm{hrs}$
(e)  $t \simeq 720~\mathrm{hrs}$.
A 7-second animation of this figure is available online (\href{https://www.predsci.com/corona/tdc/animations/RL_2023_Fig4.mp4}{www.predsci.com/corona/tdc/animations/RL\_2023\_Fig4.mp4}), showing both visualization styles for the duration of the simulation time.
}
\label{fig-aia171}
\end{figure*}

%=======================================================

\label{sec-demonstrate}
We now present a time-dependent simulation of the solar corona obtained 
with the WTD model of \citet{2018NatAs...2..913M}, driven with the
transverse electric field from Eqs.~(\ref{e:psi}-\ref{e:phi}), and 
the radial electric field from method \textsf{A}. As input to the model, we used 720 1-hour 
cadence frames of the ESFAM model \citep{2003SoPh..212..165S}, which corresponds to 30 days of evolution or just longer than one Carrington rotation. This is the same model that is implemented within the PFSS package
in SolarSoft as the SFT (surface-flux transport) module, which routinely assimilates photospheric magnetogram to best describe the surface conditions at a given time. 

For this case we use a special `fake sun' ESFAM run designed to represent a the typical conditions during solar minimum, including large-scale surface flows, the emergence and decay of bipolar patches, and the evolution of random flux across a range of scales. 
For our purposes, this type of simulation is ideal because we can sidestep some of the typical systematics of SFT maps that assimilate data.  These include: (1) zero-point offsets for the entire map because at a given assimilation step only a portion of an active region is observed; (2) unphysical evolution driven by the sweep of the assimilation window, introducing new data instantaneously, as it moves across the Sun at the solar rotation rate. We now describe
these maps in more detail.

\subsection{SFT Maps}

% EIM addition, taken from previous email chain with Marc
The main methodology for the formation of the SFT maps is described in Appendix A of \cite{2001ApJ...547..475S}, but we briefly summarize it here. The emergence pattern for bipoles is based in an automated process that randomly selects bipoles with values for their flux, location, and axial tilt based in power law distributions derived from long-term observational statistics. Once the flux has emerged---a process which is scaled to the total flux of the bipole---the polarities advect and cancel with nearby flux independently of each other. For very large bipoles (i.e., active regions), the flux is emerged as a group of smaller bipoles that aggregate to the characteristics listed above. Smaller bipoles are also emerged with similarly randomized qualities (and a correspondingly larger latitudinal distribution) in order to create a realistic background and distribution of the total flux. To process this total flux and model the cancellation on the correct time scales, there is an additional exponential flux removal term \citep[described in][]{2002ApJ...577.1006S}.

The flux in the SFT module was advected with the differential rotation profile 
taken from Table 1 of \citet{1993SoPh..145....1K}, 
in the case for the one-dimensional cross-correlation analysis applied 
to Kitt Peak magnetogram data from 1975–1991:
\begin{equation}
\Omega(\lambda) = A + B \sin^2(\lambda) + C \sin^4(\lambda).
\end{equation}
The above is a general functional form for differential rotation fits. Here,
 $\Omega$   means  the sidereal rotation rate, and $\lambda$
 is latitude. The coefficients measured by \citet{1993SoPh..145....1K}
are
\begin{eqnarray}
A&=&14.42~\mathrm{deg/day} \nonumber \\
B&=&-2.00~\mathrm{deg/day}\nonumber \\
C&=&-2.09~\mathrm{deg/day} \label{eq-difrot}
\end{eqnarray}

As implemented in the SFT model, a Carrington frame of reference is 
used, and so the solid-body sidereal  Carrington rotation rate of 
$14.18~\mathrm{deg/day}$ is subtracted off from $A$. 
Also, in the SFT model, the rotation profile does not change with time 
(i.e., no torsional oscillations are present, etc.).

The meridional flow profile $M$ is slightly more ad hoc,
 but still takes its cues from empirical measurements. Like the differential rotation profile, it is implemented in the SFT model data as a constant-in-time function and does not vary.

The SFT model's profile was originally based on the measurements 
described in \citet{1993SoPh..147..207K}, which are of the form 
$M =  f [\sin(2\theta) ]$, or more specifically, 
$\sin(2\theta) $ is the leading term in an expression also involving  
$\sin(4\theta) $. However, the current SFT model ignores those fourth order terms, and also includes a tapering function applied to the polar latitudes, so that the functional form looks is the following:

\begin{equation}
M(\theta) = v_A \sin(2\theta) g(\theta) g(\pi-\theta)
\end{equation}

Here, $M$ is the meridional flow speed and $\theta$ is colatitude, with 
the coefficient $v_A$ set to 12.7 m/s. Note that we could switch from latitude to colatitude since  
$ \sin(2\mathrm{lat.}) = \sin(2\mathrm{colat.})$.
The tapering functions $g$ only apply poleward of 40 degrees in each hemisphere:
\begin{eqnarray}
 g(\theta) & = & 1 - \exp(-a\theta^3) \nonumber \\
g(\pi-\theta) & = & 1 - \exp(-a[\pi-\theta]^3)
\end{eqnarray}

where $a = 3.0$. Without these tapering functions, the meridional flow will concentrate flux too close to the poles in the SFT model, in contrast to observations in which it is evident that the polar cap flux during solar minimum intervals is more spread out. \citet{2001ApJ...551.1099S} felt justified in using 
this tapering function given the higher uncertainties of measurements of 
meridional flows near the poles.
Note that this meridional flow profile is functionally similar to that 
given in Eq.~(3) of \citet{2001ApJ...551.1099S} but with different values 
of $v_A$ and $a$ (these values can be found in \cite{2023ApJ...946..105B}). 
\subsection{Properties of the Simulation}
The ESFAM/SFT maps were given
 as input to the MHD WTD model \citep{2018ApJ...856...75T,2018NatAs...2..913M}. 
For the  latter we used a nonuniform 
grid in $r\times\theta\times \phi$ of $269\times181\times361$ points
extending from $R_\odot$ to $30 R_\odot$.
The smallest radial grid spacing
at $r=R_\sun$   was   $\sim400$ km; the    angular resolution in $\theta$
ranged between $0.8^\circ$ at the equator and $1.7^\circ$ at the poles;
the $\phi$ mesh was uniform.
To dissipate structures that cannot be resolved
since they are  smaller than the cell size,
we prescribed a uniform resistivity $\eta$ corresponding to a
resistive diffusion time $\tau_R \sim 4\times 10^2$ hours, which
is much lower than the value in the solar corona. 
 The Alfv\'en
travel time at the base of the corona ($\tau_A=R_\sun/V_A$) for
$|\mathbf{B}|=2.205~\mathrm{G}$  and $n_0=10^8~\mathrm{cm}^{-3}$, which
are typical reference values,
is 24 minutes (Alfv\'en speed $V_A= 480~\mathrm{km/s}$),
so the Lundquist number  $\tau_R/\tau_A$ was $1\times 10^5$.
Also, in order to dissipate unresolved scales
without substantially affecting the global solution, we introduced 
a uniform
viscosity $\nu$, corresponding to a
viscous diffusion time $\tau_\nu$ such that $\tau_A/\tau_\nu=0.015$.
We prescribed fixed chromospheric values
 of density and temperature at the base of the domain of $n_0=4\times
10^{12}~\mathrm{cm^{-3}}$ and $T_0=17{,}500~\mathrm{K}$, respectively.
These values were set to form a chromospheric   ``temperature plateau''
that remains sufficiently large \citep{2009ApJ...690..902L} during
the calculation no matter how large the heating.

For the coronal heating term, we use the same WTD model parameters as the simulation described in \citet{boe21,boe22}, which is a slight update to the numbers used in  \citet{2018NatAs...2..913M}. The Poynting flux of wave energy is prescribed at the base of the corona
through an amplitude of the Els\"asser variable 
$z_0=9.63~\mathrm{km/s}$, and we set the transverse correlation scale $\lambda_0
= 0.02 R_\odot$ along with a scaling factor $B_0=8.53~\mathrm{G}$ such that
{$\lambda_\perp= \lambda_0 \sqrt{B_0/B}$} in the corona.
Similar to \citet{2018NatAs...2..913M} in adding two small
exponential heating terms to heat the low corona: 
$H_0=2.7\times 10^{-5}~\mathrm{erg/cm^3/s} $, $\lambda_0 = 0.03 R_\odot$; 
$H_0=1.6\times 10^{-8}~\mathrm{erg/cm^3/s} $, $\lambda_0 =  R_\odot$.
Likewise,  the 
wave pressure was specified from the WKB model \citep{2009ApJ...690..902L}.

We started the calculation using the first of the 720 magnetic flux maps
to calculate a potential field extrapolation. The plasma
temperature, density, and velocity were
 imposed from a 1D solar wind solution that had been calculated previously.
Then we  advanced the MHD equations for about 80 hours to relax the
system to a steady-state.
After the relaxation was accomplished,  we turned  the 
surface evolution on. The differential rotation and meridional flow 
parameters were the same as
those of Eq.~(\ref{eq-difrot}), except that,
 since our model is in the corotating
 frame of reference of the Sun, $A=0.24~\mathrm{deg/day}$ for us.

We apply method \textsf{A} to Eqs.~(\ref{e:psi}-\ref{e:phi}), using $A=1/4$ 
for Eq.~(\ref{eq:f_newdef}) and 
$D=10$ for Eq.~(\ref{eq:gfunc}),  to
drive the MHD model for a Carrington rotation.

\subsection{Coronal Evolution}

\begin{figure*}[t]
\begin{center}
\includegraphics[width=0.5\textwidth]{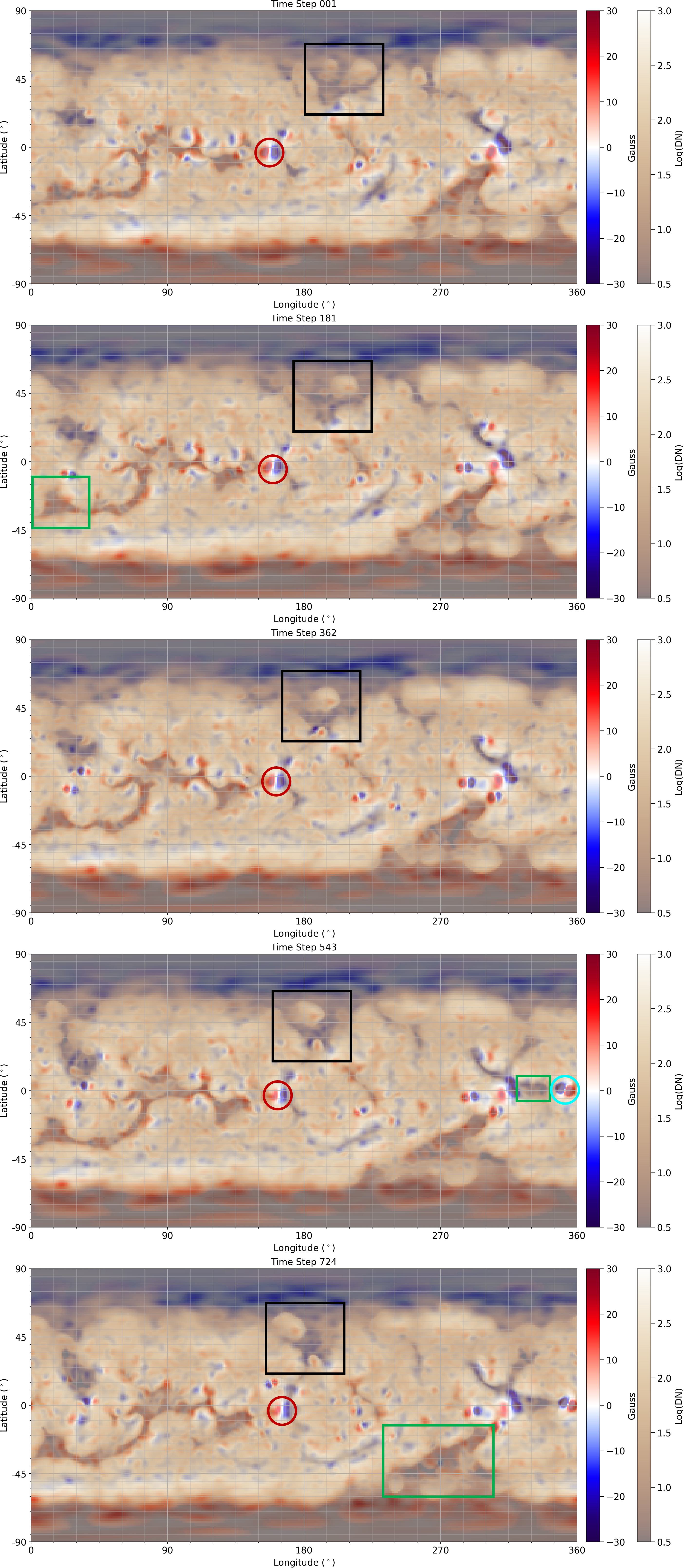}
\end{center}
\caption{Time series of time-dependent model evolution over one month, presented as latitude-longitude maps of $B_r$ overlaid with forward-modeled log-scaled synthetic SDO AIA 193 \AA\ emission from the temperature and density of the model \citep{2009ApJ...690..902L}. The black boxes indicate the same persistent open field region throughout the simulation; green boxes illustrate several other open field areas. The red circles indicate a persistent bipole, which decays during the simulation. The cyan circle shows a bipole which emerges during the course of the simulation. A 7-second animation of this figure is available online (\href{https://www.predsci.com/corona/tdc/animations/RL_2023_Fig5.mp4}{www.predsci.com/corona/tdc/animations/RL\_2023\_Fig5.mp4}), showing this map evolving over the full simulation time.}
\label{fig-br193}
\end{figure*}
 
The left-hand side of Fig.~\ref{fig-br_ch} and its associated online animation show the $B_r$ field evolution during the course of the simulation, while the right-hand side shows the open field for the same times. The evolution of $B_r$ at higher latitudes is dominated by differential rotation, and it is possible for the eye to pick up features as they are advected from west (right) to east (left), particularly in the associated animation available online. A few of the most prominent of these dynamic regions are indicated with circles on the figure. The black (left) and gold (right) rectangles appearing in all frames in the figure highlight a particularly persistent and complex region of mostly open field with a sizable embedded parasitic polarity; the whole feature drifts slowly to solar east over time due to differential rotation, and the open field in the area changes significantly during the month of simulated time. 

However, at lower latitudes, differential rotation becomes less and less discernible, while emergence and dispersion of magnetic flux become dominant. The magenta circle in the same frames shows a relatively strong bipolar region which is present at the beginning of the simulation and which undergoes decay and dissipation. Other rectangles and circles denote open field and flux emergence regions (respectively), that are more dynamic or that emerge during the course of the simulation.

The results of these time-dependent effects on simulated emission are illustrated in Figs.~\ref{fig-aia171} and \ref{fig-br193}. We use the temperature and density from the simulation to forward-model the emission in the SDO AIA 171 \AA\ channel for this figure. The left-hand column of Fig.~\ref{fig-aia171} shows a 3D projection observed from an Earth-like point of view at the same times as the preceding figures (i.e. the longitude of the observer changes with the Carrington rotation rate). To better understand the evolution of the emission, we can compute Carrington latitude-longitude maps of forward modeled AIA emission by integrating along the radial direction for all points on the model. These are shown in the right column of Fig.~\ref{fig-aia171}, with the same annotations as the preceding figure to aid in by-eye comparisons. Here, the bipoles appear as bright active regions, while particularly bright patches within the active regions which evolve rapidly (best seen in the associated animation) are signatures of thermal nonequilibrium within these areas of highly-stratified heating. The open fields contain somewhat lower emission than the closed field, though not as dark as would be expected from a higher-temperature emission line (such as 193 \AA, shown in the following figure).

In Fig.~\ref{fig-br193} we show of the surface $B_r$ prescribed at the lower boundary, overlaid with forward-modeled SDO AIA 193 \AA\ emission. This composite presentation shows the close relationship between the changing magnetic boundary conditions and the coronal evolution. Features visible in emission in the polar coronal holes are matched by the unipolar signatures in the underlying $B_r$ map, and are likewise advected by differential rotation within the coronal holes. The black box repeated on each panel of Fig.~\ref{fig-br193} highlights one such persistent open flux region. If only differential rotation were present, the open flux evolution would be (mostly) rigid, as observations  \citep{1975SoPh...42..135T} show, and both potential \citep{1996Sci...271..464W} and MHD  \citep{2005ApJ...625..463L}  models indicate. However, magnetic flux evolution contributes to continuous changes in the boundary between open and closed-field region. This is particularly evident (but not limited to) the equatorial region between $270^\circ$ and $360^\circ$ longitude. Here on frame 543, flux emergence and open-field evolution are highlighted with the green box and cyan circle, which highlight an area in which emerging flux results in an expanded neighboring open field region. In contrast, the red circle that appears in each panel illustrates a case of magnetic flux decay, as a bipolar region becomes more diffuse over the month timeframe of the simulation. Because of the richness of the EUV evolution and its relationship to interchange reconnection with open and closed fields, we explore the effects of these and related coronal signatures in the companion paper analyzing this simulation \citep{mason2023timedependent}.

\subsection{Dipped Field Lines}
\begin{figure*}
\begin{center}
\includegraphics[width=\textwidth]{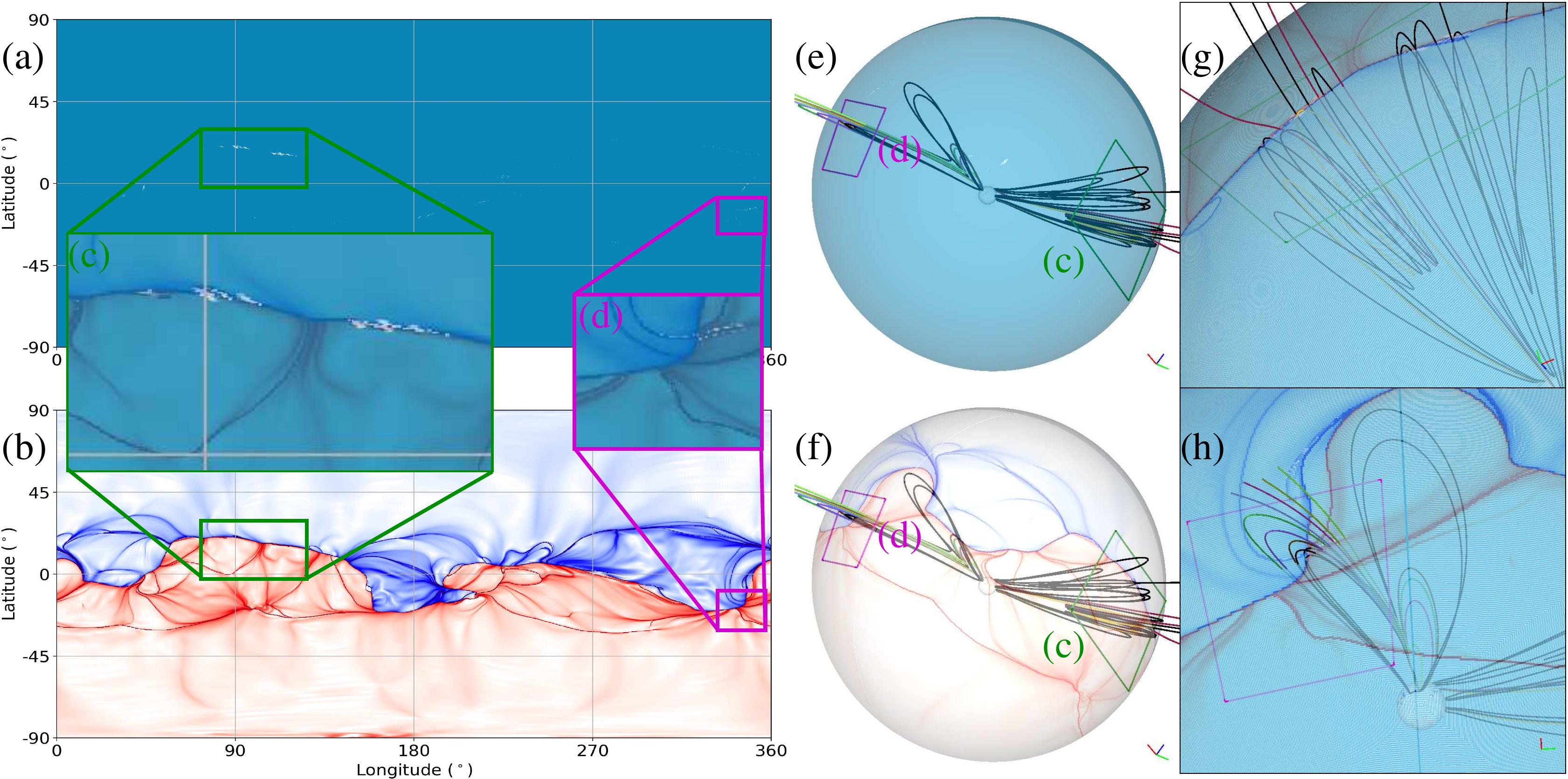} 
\end{center}
\caption{Dipped magnetic field lines (field lines that have a reversal of the
sign of $B_r$) at $t= 524~\mathrm{hrs}$. \textbf{(a)} Footpoints of dipped field
lines at $r=19 R_\odot$ as latitude-longitude map. \textbf{(b)} Latitude-longitude map 
of $\slog Q$ at the same height. \textbf{(c)} Enlargement and superposition 
of a region of (a) and (b) where the dipped field-lines are
associated with the current sheet. \textbf{(d)} Enlargement and superposition
of another region of (a) and (b) where the dipped field lines are
associated with a single-polarity separatrix line. \textbf{(e)} The 
 $r=19 R_\odot$ semi-transparent surface colored as in (a) with dipped
field-lines footpoints and
marked with the (c) and (d) regions.  Representative field
lines are visible and the solar surface is at
the center.  \textbf{(f)} The same as (e) but colored
as in (b) with $\slog Q$.
\textbf{(g)} 3D enlargement of the (c) region with magnetic field lines.
The solar surface is at the bottom right.
\textbf{(h)}  3D enlargement of the (d) region with magnetic field lines.
}
\label{fig-dips_slog1}
\end{figure*}
Parker Solar Probe, during its first perihelion, discovered that the
radial magnetic field  was continuously interrupted by switchbacks
on a time scale of less than a
second to more than one hour \citep{2020ApJS..246...39D}.
Although our simulation was aimed at reproducing lower frequency phenomena and
at larger length scales, we also investigated the formation of  dips in magnetic
field lines. First, for each cell in the computational domain we determined
whether there was a
dip in its neighborhood by tracing a magnetic field segment and checking
whether $B_r$ changed sign along it. Then, from each point on an
$r=19 R_\odot$  surface,
 we traced a field line and verified whether
it passed through a cell having a dip. The resulting map is shown in
Fig.~\ref{fig-dips_slog1}a, where the seed points of
dipped field lines are painted in white. Figure ~\ref{fig-dips_slog1}b has
a map of $\slog Q$ \citep[signed logarithm of the squashing factor 
$Q$, ][]{2007ApJ...660..863T} on the same surface. At low latitude, 
the so-called S-web generally coincides with the region where the slow solar wind 
is observed \citep{2011ApJ...731..112A}. A comparison between
panel (a) and (b) indicates that dipped field lines have connections with
 the S-web. In particular, we enlarged two regions in the S-web map,
the corresponding areas of the dipped field-lines map, 
and superimposed them in panels (c) and (d). 
While the 
field lines associated with reversal in the sign of $B_r$ lie along
high $Q$ lines, only some have footpoints aligned with the current sheet 
(Fig.~\ref{fig-dips_slog1}c). Other dipped field-lines cross single-polarity
separatrices as in Figs.~\ref{fig-dips_slog1}d.
In panel (e),
the dipped field-line map of panel (a) is shown as a semi-transparent 
surface with two
groups of
representative field lines intersecting the areas of panels (c) and (d).
Likewise, panel (f) presents the equivalent point of view for the
$\slog Q$ map. Panels (g) and (h) present 3D enlargements  around 
the (c) and (d) areas respectively. Some of the field lines in (g), which 
are all associated with the current sheet, are arranged in a flux-rope.
On the other hand, some of the field lines in panel (h), which 
are all associated with a same-polarity high $Q$ line, form a
  \textsf{V} shape, indicative of interchange reconnection (e.g.,
see  Fig.~5c and 6c of \citet{2005ApJ...625..463L}).

\subsection{Charge-States in the Static and Time Dependent Corona}
\begin{figure*}
\begin{center}
\includegraphics[width=.8 \textwidth]{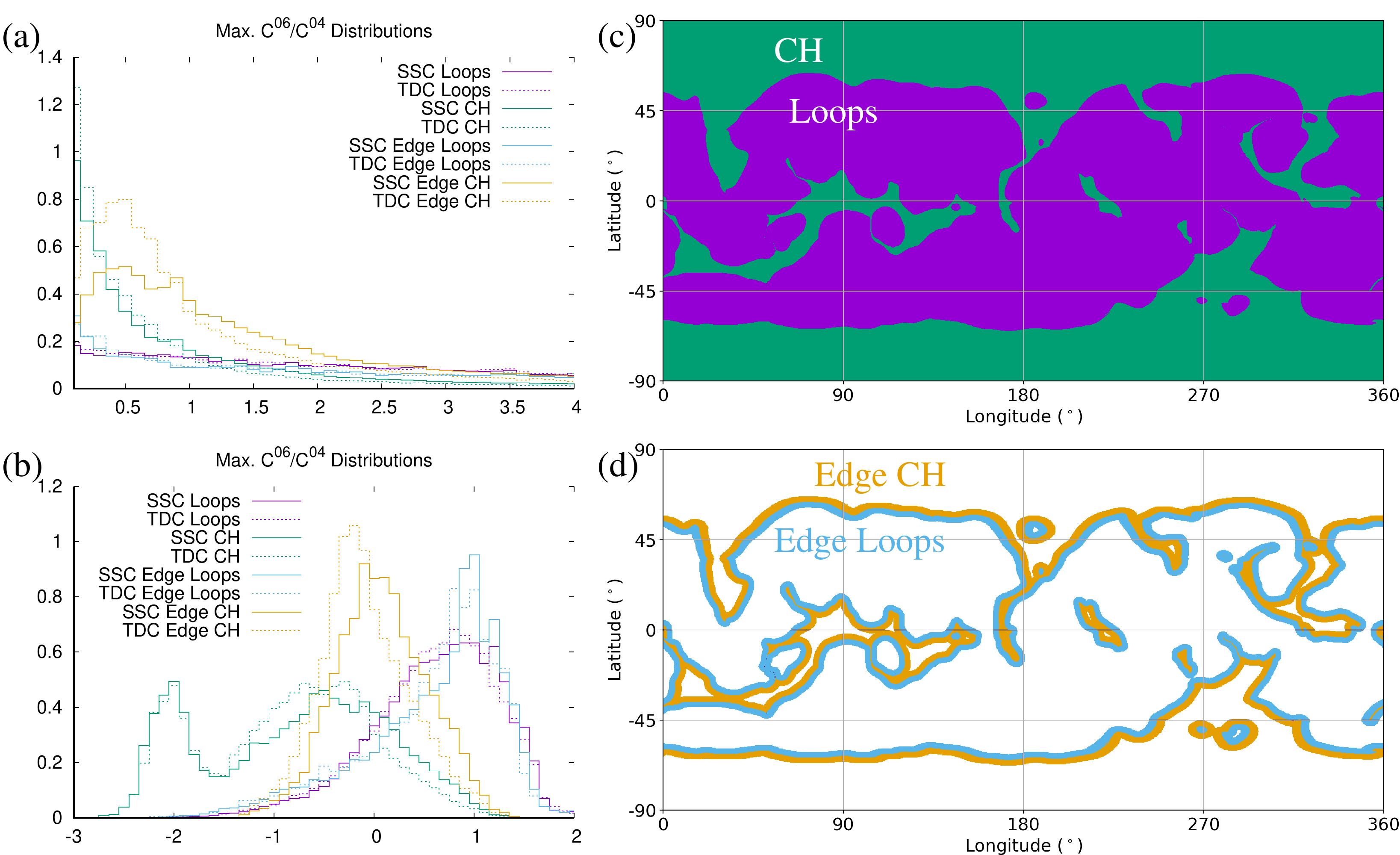}
\end{center}
\caption{\textbf{(a)} Statistical distributions of the maximum value of the 
$C^{06}/C^{04}$ ratio evaluated along magnetic field lines in the time-dependent corona and
steady-state corona models. We show distributions for  loops (i.e., closed-field regions), coronal holes 
(open-field regions),
loops at the edge of open-field areas, and edges of coronal holes.
\textbf{(b)} The same as (a) with a logarithmic scale for the axes.
\textbf{(c)} The areas over which the loops (violet) and coronal-holes (green) 
distributions
are defined. \textbf{(d)} The  same as (c) for  the edge loops (cyan) and
edge coronal-holes (gold) distributions.
}
\label{fig-qfe}
\end{figure*}

To understand the effects of the continuous evolution of the 
surface magnetic fields on the ion charge-state distributions
of the corona and solar wind, we compared the results of the
time-dependent model at $t=524~\mathrm{hrs} $   with its
corresponding steady-state model. This steady-state-corona (SSC) model was obtained
by  stopping all surface
flows and magnetic flux evolution and letting the system relax for 
approximately 80 hours.
We calculated  the maxima of the $C^{06}/C^{04}$ ratio along magnetic
field lines.
This is the ratio recommended by \citet{2015ApJ...812L..28L} for analysis and
comparison between models and \emph{in situ} data as less sensitive to 
photoionization. In Fig.~\ref{fig-qfe} we show statistical    
distributions using either a linear (a) or a logarithmic scale (b) in 
the  $x$ axis.  The solid lines are associated with the SSC model, the
dotted lines with the TDC one. We distinguish four regions: loops (closed-field regions),
coronal holes (open-field regions), which are mapped in panel (c), edge loops (i.e., long loops bordering open-field regions),
edge corona holes (i.e., open-field lines but close to closed-field areas), 
which appear in panel (d).  Although the loops (violet) distributions
are similar, 
we notice that the TDC model has a  $C^{06}/C^{04}$   distribution in
  coronal holes (green) that is
significantly higher for smaller ratios than that of the SSC model. This is even more
evident in the distribution of field lines at the edge of coronal holes (gold), 
with visibly higher distribution values for  $C^{06}/C^{04} \lesssim 1$.

%%%%%%%%%%%%%%%%%%%%%%
%%%%%%%%%%%%%%%%%%%%%%
\section{DISCUSSION AND CONCLUSION}
\label{sec-dis}

We applied to our MHD model
 a technique that drives  the evolution of the 
photospheric magnetic flux and of the  surface 
flows to calculate the response of the solar corona. 
 The electric field, which the 
technique provides, may be corrected with two methods to reduce 
the formation of current boundary layers. We tested the two corrective
 methods by
comparing results with a simplified, reference simulation and found that
either gives results in satisfactory agreement. We then used our 
technique with method \textsf{A}
 to calculate the evolution of the corona for a whole month,
 using a sequence of balanced
magnetic-flux maps obtained with the SFT module.
From the simulation data, we produced emission images that show a
continuous reconfiguration of the corona as active regions emerge and
disperse and surface flows rearrange the flux. 
Dipped field lines that are formed during the
computation appear to be associated with the S-web. In particular, we
identified a fluxrope in the current sheet, while field lines of 
single-polarity areas show the typical  pattern of interchange reconnection.
Ions fractional charge-states were evolved alongside the dynamics. 
The distribution of the $C^{06}/C^{04}$ ion charge-state ratio appear to differ
between the time-dependent model and the steady-state corona. 
This difference is visible in the open-field regions, for which the
time-dependent model has a distribution more skewed to lower charge states than
the steady-state corona, and  particularly evident in the field
lines close to the border of coronal holes.

%\cd{RL read: I re-worked the concluding paragraph(s) to better capture what I think is going on with the charge-states and then provide a higher level discussion of the importance and future of the method. Please check.}

Counterintuitively, it is the TDC model that has the lower ratios. The conventional wisdom for enhanced ratios in the slow-wind is that they represent hotter plasma from closed loops that has been liberated through interchange reconnection. In the TDC model this can occur either by surface flows and/or evolution or by a thermal and/or tearing instability introducing reconnection at the streamer cusp itself \citep[e.g.][]{reville20a}, while in the SSC model this can only occur through the latter processes. On the other hand, the TDC model is slightly more open and the streamer loops naturally have a shorter lifetime as the surface flux evolves and their geometry changes in time, leading to a lower temperature on average. It seems that in this case, with relatively modest resolution and minimal energization in the transverse field (\S\ref{ss:corrections}) that the latter process wins out. In future work we aim to study how additional shear (helicity) emerged at global scales will influence the charge-state ratios as well as the role of small-scale flux patches that inducing interchange reconnection near the coronal base as they evolve \citep[e.g.][]{sterling20,bale22_arxiv}.

Ultimately we have demonstrated a practical technique for time-dependently driving a global coronal model from only a sequence of magnetic maps. This leads to an inherent time-evolution of features in the model corona that is not possible to capture with traditional steady-state methods. Such a capability is crucial for answering long-standing questions about how closed and open fields evolve in the corona, and how the field and plasma properties of the observed slow-solar wind are formed. Lastly, with the advent of modern SFT models that readily assimilate observations to produce time-sequences of full-sun magnetic maps, we expect simple but robust driving techniques like those introduced here to play a key role in the next generation of solar wind and CME models.

This work was supported by the NASA Heliophysics Living With a Star Science and Strategic Capabilities programs (grant numbers 80NSSC20K0192, 80NSSC22K1021, and 80NSSC22K0893), the NASA Heliophysics System Observatory Connect program (grant number 80NSSC20K1285), and the NSF PREEVENTS program (grant ICER1854790). Computing resources supporting this work were provided by the NASA High-End Computing (HEC) Program through the NASA Advanced Supercomputing (NAS) Division at Ames Research Center and by the Expanse supercomputer at the San Diego Supercomputing Center through the NSF ACCESS and XSEDE programs. MLD acknowledges support from NASA under contract NNG04EA00C as part of the AIA science team.

\bibliographystyle{aasjournal}
\bibliography{mybib,cooper}

\begin{thebibliography}{}
\expandafter\ifx\csname natexlab\endcsname\relax\def\natexlab#1{#1}\fi
\providecommand{\url}[1]{\href{#1}{#1}}
\providecommand{\dodoi}[1]{doi:~\href{http://doi.org/#1}{\nolinkurl{#1}}}
\providecommand{\doeprint}[1]{\href{http://ascl.net/#1}{\nolinkurl{http://ascl.net/#1}}}
\providecommand{\doarXiv}[1]{\href{https://arxiv.org/abs/#1}{\nolinkurl{https://arxiv.org/abs/#1}}}

\bibitem[{{Antiochos} {et~al.}(2011){Antiochos}, {Miki{\'c}}, {Titov},
  {Lionello}, \& {Linker}}]{2011ApJ...731..112A}
{Antiochos}, S.~K., {Miki{\'c}}, Z., {Titov}, V.~S., {Lionello}, R., \&
  {Linker}, J.~A. 2011, \apj, 731, 112, \dodoi{10.1088/0004-637X/731/2/112}

\bibitem[{{Arge} {et~al.}(2010){Arge}, {Henney}, {Koller}, {Compeau}, {Young},
  {MacKenzie}, {Fay}, \& {Harvey}}]{2010AIPC.1216..343A}
{Arge}, C.~N., {Henney}, C.~J., {Koller}, J., {et~al.} 2010, Twelfth
  International Solar Wind Conference, 1216, 343, \dodoi{10.1063/1.3395870}

\bibitem[{{Bale} {et~al.}(2022){Bale}, {Drake}, {McManus}, {Desai}, {Badman},
  {Larson}, {Swisdak}, {Horbury}, {Raouafi}, {Phan}, {Velli}, {McComas},
  {Cohen}, {Mitchell}, {Panasenco}, \& {Kasper}}]{bale22_arxiv}
{Bale}, S.~D., {Drake}, J.~F., {McManus}, M.~D., {et~al.} 2022, arXiv e-prints,
  arXiv:2208.07932.
\newblock \doarXiv{2208.07932}

\bibitem[{{Barnes} {et~al.}(2023){Barnes}, {DeRosa}, {Jones}, {Arge}, {Henney},
  \& {Cheung}}]{2023ApJ...946..105B}
{Barnes}, G., {DeRosa}, M.~L., {Jones}, S.~I., {et~al.} 2023, \apj, 946, 105,
  \dodoi{10.3847/1538-4357/acba8e}

\bibitem[{{Benz}(2017)}]{Benz2017}
{Benz}, A.~O. 2017, Living Reviews in Solar Physics, 14, 2,
  \dodoi{10.1007/s41116-016-0004-3}

\bibitem[{{Boe} {et~al.}(2021){Boe}, {Habbal}, {Downs}, \&
  {Druckm{\"u}ller}}]{boe21}
{Boe}, B., {Habbal}, S., {Downs}, C., \& {Druckm{\"u}ller}, M. 2021, \apj, 912,
  44, \dodoi{10.3847/1538-4357/abea79}

\bibitem[{{Boe} {et~al.}(2022){Boe}, {Habbal}, {Downs}, \&
  {Druckm{\"u}ller}}]{boe22}
---. 2022, \apj, 935, 173, \dodoi{10.3847/1538-4357/ac8101}

\bibitem[{{Breech} {et~al.}(2008){Breech}, {Matthaeus}, {Minnie}, {Bieber},
  {Oughton}, {Smith}, \& {Isenberg}}]{2008JGRA..113.8105B}
{Breech}, B., {Matthaeus}, W.~H., {Minnie}, J., {et~al.} 2008, Journal of
  Geophysical Research (Space Physics), 113, 8105, \dodoi{10.1029/2007JA012711}

\bibitem[{{Chandran} \& {Hollweg}(2009)}]{2009ApJ...707.1659C}
{Chandran}, B. D.~G., \& {Hollweg}, J.~V. 2009, \apj, 707, 1659,
  \dodoi{10.1088/0004-637X/707/2/1659}

\bibitem[{{Cheung} \& {DeRosa}(2012)}]{cheung12}
{Cheung}, M.~C.~M., \& {DeRosa}, M.~L. 2012, \apj, 757, 147,
  \dodoi{10.1088/0004-637X/757/2/147}

\bibitem[{{Cranmer} {et~al.}(2007){Cranmer}, {van Ballegooijen}, \&
  {Edgar}}]{2007ApJS..171..520C}
{Cranmer}, S.~R., {van Ballegooijen}, A.~A., \& {Edgar}, R.~J. 2007, \apjs,
  171, 520, \dodoi{10.1086/518001}

\bibitem[{{Dere} {et~al.}(1997){Dere}, {Landi}, {Mason}, {Monsignori Fossi}, \&
  {Young}}]{1997A&AS..125..149D}
{Dere}, K.~P., {Landi}, E., {Mason}, H.~E., {Monsignori Fossi}, B.~C., \&
  {Young}, P.~R. 1997, \aaps, 125, 149

\bibitem[{{Dmitruk} {et~al.}(2001){Dmitruk}, {Milano}, \&
  {Matthaeus}}]{2001ApJ...548..482D}
{Dmitruk}, P., {Milano}, L.~J., \& {Matthaeus}, W.~H. 2001, \apj, 548, 482,
  \dodoi{10.1086/318685}

\bibitem[{{Downs} {et~al.}(2013){Downs}, {Linker}, {Miki{\'c}}, {Riley},
  {Schrijver}, \& {Saint-Hilaire}}]{2013Sci...340.1196D}
{Downs}, C., {Linker}, J.~A., {Miki{\'c}}, Z., {et~al.} 2013, Science, 340,
  1196, \dodoi{10.1126/science.1236550}

\bibitem[{{Dudok de Wit} {et~al.}(2020){Dudok de Wit}, {Krasnoselskikh},
  {Bale}, {Bonnell}, {Bowen}, {Chen}, {Froment}, {Goetz}, {Harvey},
  {Jagarlamudi}, {Larosa}, {MacDowall}, {Malaspina}, {Matthaeus}, {Pulupa},
  {Velli}, \& {Whittlesey}}]{2020ApJS..246...39D}
{Dudok de Wit}, T., {Krasnoselskikh}, V.~V., {Bale}, S.~D., {et~al.} 2020,
  \apjs, 246, 39, \dodoi{10.3847/1538-4365/ab5853}

\bibitem[{{Fisher} {et~al.}(2010){Fisher}, {Welsch}, {Abbett}, \&
  {Bercik}}]{fisher10}
{Fisher}, G.~H., {Welsch}, B.~T., {Abbett}, W.~P., \& {Bercik}, D.~J. 2010,
  \apj, 715, 242, \dodoi{10.1088/0004-637X/715/1/242}

\bibitem[{{Fisk} \& {Kasper}(2020)}]{fisk20}
{Fisk}, L.~A., \& {Kasper}, J.~C. 2020, \apjl, 894, L4,
  \dodoi{10.3847/2041-8213/ab8acd}

\bibitem[{{Fritsch} \& {Carlson}(1980)}]{1980SJNA...17..238F}
{Fritsch}, F.~N., \& {Carlson}, R.~E. 1980, SIAM Journal on Numerical Analysis,
  17, 238, \dodoi{10.1137/0717021}

\bibitem[{{Heinemann} \& {Olbert}(1980)}]{1980JGR....85.1311H}
{Heinemann}, M., \& {Olbert}, S. 1980, \jgr, 85, 1311,
  \dodoi{10.1029/JA085iA03p01311}

\bibitem[{Higginson {et~al.}(2017)Higginson, Antiochos, DeVore, Wyper, \&
  Zurbuchen}]{Higginson2017}
Higginson, A.~K., Antiochos, S.~K., DeVore, C.~R., Wyper, P.~F., \& Zurbuchen,
  T.~H. 2017, The Astrophysical Journal, 840, L10,
  \dodoi{10.3847/2041-8213/aa6d72}

\bibitem[{{Jin} {et~al.}(2012){Jin}, {Manchester}, {van der Holst},
  {Gruesbeck}, {Frazin}, {Landi}, {Vasquez}, {Lamy}, {Llebaria}, {Fedorov},
  {Toth}, \& {Gombosi}}]{2012ApJ...745....6J}
{Jin}, M., {Manchester}, W.~B., {van der Holst}, B., {et~al.} 2012, \apj, 745,
  6, \dodoi{10.1088/0004-637X/745/1/6}

\bibitem[{{Kepko} {et~al.}(2016){Kepko}, {Viall}, {Antiochos}, {Lepri},
  {Kasper}, \& {Weberg}}]{kepko16}
{Kepko}, L., {Viall}, N.~M., {Antiochos}, S.~K., {et~al.} 2016, \grl, 43, 4089,
  \dodoi{10.1002/2016GL068607}

\bibitem[{{Komm} {et~al.}(1993{\natexlab{a}}){Komm}, {Howard}, \&
  {Harvey}}]{1993SoPh..145....1K}
{Komm}, R.~W., {Howard}, R.~F., \& {Harvey}, J.~W. 1993{\natexlab{a}},
  \solphys, 145, 1, \dodoi{10.1007/BF00627979}

\bibitem[{{Komm} {et~al.}(1993{\natexlab{b}}){Komm}, {Howard}, \&
  {Harvey}}]{1993SoPh..147..207K}
---. 1993{\natexlab{b}}, \solphys, 147, 207, \dodoi{10.1007/BF00690713}

\bibitem[{{Landi} \& {Lepri}(2015)}]{2015ApJ...812L..28L}
{Landi}, E., \& {Lepri}, S.~T. 2015, \apjl, 812, L28,
  \dodoi{10.1088/2041-8205/812/2/L28}

\bibitem[{{Landi} {et~al.}(2013){Landi}, {Young}, {Dere}, {Del Zanna}, \&
  {Mason}}]{2013ApJ...763...86L}
{Landi}, E., {Young}, P.~R., {Dere}, K.~P., {Del Zanna}, G., \& {Mason}, H.~E.
  2013, \apj, 763, 86, \dodoi{10.1088/0004-637X/763/2/86}

\bibitem[{{Linker} {et~al.}(2011){Linker}, {Lionello}, {Miki{\'c}}, {Titov}, \&
  {Antiochos}}]{2011ApJ...731..110L}
{Linker}, J.~A., {Lionello}, R., {Miki{\'c}}, Z., {Titov}, V.~S., \&
  {Antiochos}, S.~K. 2011, \apj, 731, 110, \dodoi{10.1088/0004-637X/731/2/110}

\bibitem[{{Linker} {et~al.}(2003){Linker}, {Miki{\'c}}, {Lionello}, {Riley},
  {Amari}, \& {Odstrcil}}]{2003PhPl...10.1971L}
{Linker}, J.~A., {Miki{\'c}}, Z., {Lionello}, R., {et~al.} 2003, Physics of
  Plasmas, 10, 1971

\bibitem[{{Linker} {et~al.}(1999){Linker}, {Miki{\'c}}, {Biesecker}, {Forsyth},
  {Gibson}, {Lazarus}, {Lecinski}, {Riley}, {Szabo}, \&
  {Thompson}}]{1999JGR...104.9809L}
{Linker}, J.~A., {Miki{\'c}}, Z., {Biesecker}, D.~A., {et~al.} 1999, \jgr, 104,
  9809.
\newblock
  \url{http://adsabs.harvard.edu/cgi-bin/nph-bib_query?bibcode=1999JGR...104.9809L&db_key=AST}

\bibitem[{{Lionello} {et~al.}(2019){Lionello}, {Downs}, {Linker}, {Miki{\'c}},
  {Raymond}, {Shen}, \& {Velli}}]{2019SoPh..294...13L}
{Lionello}, R., {Downs}, C., {Linker}, J.~A., {et~al.} 2019, \solphys, 294, 13,
  \dodoi{10.1007/s11207-019-1401-2}

\bibitem[{{Lionello} {et~al.}(2013){Lionello}, {Downs}, {Linker},
  {T{\"o}r{\"o}k}, {Riley}, \& {Miki{\'c}}}]{2013ApJ...777...76L}
---. 2013, \apj, 777, 76, \dodoi{10.1088/0004-637X/777/1/76}

\bibitem[{{Lionello} {et~al.}(2009){Lionello}, {Linker}, \&
  {Miki{\'c}}}]{2009ApJ...690..902L}
{Lionello}, R., {Linker}, J.~A., \& {Miki{\'c}}, Z. 2009, \apj, 690, 902,
  \dodoi{10.1088/0004-637X/690/1/902}

\bibitem[{{Lionello} {et~al.}(2006){Lionello}, {Linker}, {Miki{\'c}}, \&
  {Riley}}]{2006ApJ...642L..69L}
{Lionello}, R., {Linker}, J.~A., {Miki{\'c}}, Z., \& {Riley}, P. 2006, \apjl,
  642, L69, \dodoi{10.1086/504289}

\bibitem[{{Lionello} {et~al.}(2005){Lionello}, {Riley}, {Linker}, \&
  {Miki{\'c}}}]{2005ApJ...625..463L}
{Lionello}, R., {Riley}, P., {Linker}, J.~A., \& {Miki{\'c}}, Z. 2005, \apj,
  625, 463, \dodoi{10.1086/429268}

\bibitem[{{Lionello} {et~al.}(2020){Lionello}, {Titov}, {Miki{\'c}}, \&
  {Linker}}]{2020ApJ...891...14L}
{Lionello}, R., {Titov}, V.~S., {Miki{\'c}}, Z., \& {Linker}, J.~A. 2020, \apj,
  891, 14, \dodoi{10.3847/1538-4357/ab68d9}

\bibitem[{{Luhmann} {et~al.}(2022){Luhmann}, {Li}, {Lee}, {Jian}, {Arge}, \&
  {Riley}}]{luhmann22}
{Luhmann}, J.~G., {Li}, Y., {Lee}, C.~O., {et~al.} 2022, Space Weather, 20,
  e2022SW003110, \dodoi{10.1029/2022SW003110}

\bibitem[{{Lumme} {et~al.}(2017){Lumme}, {Pomoell}, \& {Kilpua}}]{lumme17}
{Lumme}, E., {Pomoell}, J., \& {Kilpua}, E.~K.~J. 2017, \solphys, 292, 191,
  \dodoi{10.1007/s11207-017-1214-0}

\bibitem[{Mason {et~al.}(2023)Mason, Lionello, Downs, Linker, \&
  Caplan}]{mason2023timedependent}
Mason, E.~I., Lionello, R., Downs, C., Linker, J.~A., \& Caplan, R.~M. 2023,
  Time-Dependent Dynamics of the Corona. Submitted to \apj.
\newblock \doarXiv{2306.11956}

\bibitem[{{Matthaeus} {et~al.}(1999){Matthaeus}, {Zank}, {Oughton}, {Mullan},
  \& {Dmitruk}}]{1999ApJ...523L..93M}
{Matthaeus}, W.~H., {Zank}, G.~P., {Oughton}, S., {Mullan}, D.~J., \&
  {Dmitruk}, P. 1999, \apjl, 523, L93, \dodoi{10.1086/312259}

\bibitem[{{Miki\'c} {et~al.}(1999){Miki\'c}, {Linker}, {Schnack}, {Lionello},
  \& {Tarditi}}]{1999PhPl....6.2217M}
{Miki\'c}, Z., {Linker}, J.~A., {Schnack}, D.~D., {Lionello}, R., \& {Tarditi},
  A. 1999, Phys. of Plasmas, 6, 2217.
\newblock
  \url{http://adsabs.harvard.edu/cgi-bin/nph-bib_query?bibcode=1999PhPl....6.2217M&db_key=PHY}

\bibitem[{{Miki{\'c}} {et~al.}(2018){Miki{\'c}}, {Downs}, {Linker}, {Caplan},
  {Mackay}, {Upton}, {Riley}, {Lionello}, {T{\"o}r{\"o}k}, {Titov}, {Wijaya},
  {Druckm{\"u}ller}, {Pasachoff}, \& {Carlos}}]{2018NatAs...2..913M}
{Miki{\'c}}, Z., {Downs}, C., {Linker}, J.~A., {et~al.} 2018, Nature Astronomy,
  2, 913, \dodoi{10.1038/s41550-018-0562-5}

\bibitem[{{Odstrcil} {et~al.}(2020){Odstrcil}, {Mays}, {Hess}, {Jones},
  {Henney}, \& {Arge}}]{odstrcil20}
{Odstrcil}, D., {Mays}, M.~L., {Hess}, P., {et~al.} 2020, \apjs, 246, 73,
  \dodoi{10.3847/1538-4365/ab77cb}

\bibitem[{{Oran} {et~al.}(2015){Oran}, {Landi}, {van der Holst}, {Lepri},
  {V{\'a}squez}, {Nuevo}, {Frazin}, {Manchester}, {Sokolov}, \&
  {Gombosi}}]{2015ApJ...806...55O}
{Oran}, R., {Landi}, E., {van der Holst}, B., {et~al.} 2015, \apj, 806, 55,
  \dodoi{10.1088/0004-637X/806/1/55}

\bibitem[{Raouafi {et~al.}(2016)Raouafi, Patsourakos, Pariat, Young, Sterling,
  Savcheva, Shimojo, Moreno-Insertis, DeVore, Archontis, T{\"{o}}r{\"{o}}k,
  Mason, Curdt, Meyer, Dalmasse, \& Matsui}]{Raouafi2016a}
Raouafi, N.~E., Patsourakos, S., Pariat, E., {et~al.} 2016, Space Science
  Reviews, 201, 1, \dodoi{10.1007/s11214-016-0260-5}

\bibitem[{{R{\'e}ville} {et~al.}(2020){R{\'e}ville}, {Velli}, {Rouillard},
  {Lavraud}, {Tenerani}, {Shi}, \& {Strugarek}}]{reville20a}
{R{\'e}ville}, V., {Velli}, M., {Rouillard}, A.~P., {et~al.} 2020, \apjl, 895,
  L20, \dodoi{10.3847/2041-8213/ab911d}

\bibitem[{{Riley} {et~al.}(2006){Riley}, {Linker}, {Miki{\'c}}, {Lionello},
  {Ledvina}, \& {Luhmann}}]{riley06_pfss}
{Riley}, P., {Linker}, J.~A., {Miki{\'c}}, Z., {et~al.} 2006, \apj, 653, 1510,
  \dodoi{10.1086/508565}

\bibitem[{{Schnack} {et~al.}(1987){Schnack}, {Barnes}, {Mikic}, {Harned}, \&
  {Caramana}}]{1987JCoPh..70..330S}
{Schnack}, D.~D., {Barnes}, D.~C., {Mikic}, Z., {Harned}, D.~S., \& {Caramana},
  E.~J. 1987, Journal of Computational Physics, 70, 330,
  \dodoi{10.1016/0021-9991(87)90186-0}

\bibitem[{{Schrijver}(2001)}]{2001ApJ...547..475S}
{Schrijver}, C.~J. 2001, \apj, 547, 475, \dodoi{10.1086/318333}

\bibitem[{{Schrijver} {et~al.}(2002){Schrijver}, {De Rosa}, \&
  {Title}}]{2002ApJ...577.1006S}
{Schrijver}, C.~J., {De Rosa}, M.~L., \& {Title}, A.~M. 2002, \apj, 577, 1006,
  \dodoi{10.1086/342247}

\bibitem[{{Schrijver} \& {DeRosa}(2003)}]{2003SoPh..212..165S}
{Schrijver}, C.~J., \& {DeRosa}, M.~L. 2003, \solphys, 212, 165

\bibitem[{{Schrijver} \& {Title}(2001)}]{2001ApJ...551.1099S}
{Schrijver}, C.~J., \& {Title}, A.~M. 2001, \apj, 551, 1099,
  \dodoi{10.1086/320237}

\bibitem[{Scott {et~al.}(2021)Scott, Pontin, Antiochos, DeVore, \&
  Wyper}]{Scott2021}
Scott, R.~B., Pontin, D.~I., Antiochos, S.~K., DeVore, C.~R., \& Wyper, P.~F.
  2021, The Astrophysical Journal, 913, 64, \dodoi{10.3847/1538-4357/abec4f}

\bibitem[{{Shen} {et~al.}(2015){Shen}, {Raymond}, {Murphy}, \&
  {Lin}}]{2015A&C....12....1S}
{Shen}, C., {Raymond}, J.~C., {Murphy}, N.~A., \& {Lin}, J. 2015, Astronomy and
  Computing, 12, 1, \dodoi{10.1016/j.ascom.2015.04.003}

\bibitem[{{Sokolov} {et~al.}(2013){Sokolov}, {van der Holst}, {Oran}, {Downs},
  {Roussev}, {Jin}, {Manchester}, {Evans}, \& {Gombosi}}]{2013ApJ...764...23S}
{Sokolov}, I.~V., {van der Holst}, B., {Oran}, R., {et~al.} 2013, \apj, 764,
  23, \dodoi{10.1088/0004-637X/764/1/23}

\bibitem[{{Solanki} {et~al.}(2020){Solanki}, {del Toro Iniesta}, {Woch},
  {Gandorfer}, {Hirzberger}, {Alvarez-Herrero}, {Appourchaux}, {Mart{\'\i}nez
  Pillet}, {P{\'e}rez-Grande}, {Sanchis Kilders}, {Schmidt}, {G{\'o}mez Cama},
  {Michalik}, {Deutsch}, {Fernandez-Rico}, {Grauf}, {Gizon}, {Heerlein},
  {Kolleck}, {Lagg}, {Meller}, {M{\"u}ller}, {Sch{\"u}hle}, {Staub}, {Albert},
  {Alvarez Copano}, {Beckmann}, {Bischoff}, {Busse}, {Enge}, {Frahm},
  {Germerott}, {Guerrero}, {L{\"o}ptien}, {Meierdierks}, {Oberdorfer},
  {Papagiannaki}, {Ramanath}, {Schou}, {Werner}, {Yang}, {Zerr}, {Bergmann},
  {Bochmann}, {Heinrichs}, {Meyer}, {Monecke}, {M{\"u}ller}, {Sperling},
  {{\'A}lvarez Garc{\'\i}a}, {Aparicio}, {Balaguer Jim{\'e}nez}, {Bellot
  Rubio}, {Cobos Carracosa}, {Girela}, {Hern{\'a}ndez Exp{\'o}sito}, {Herranz},
  {Labrousse}, {L{\'o}pez Jim{\'e}nez}, {Orozco Su{\'a}rez}, {Ramos},
  {Barandiar{\'a}n}, {Bastide}, {Campuzano}, {Cebollero}, {D{\'a}vila},
  {Fern{\'a}ndez-Medina}, {Garc{\'\i}a Parejo}, {Garranzo-Garc{\'\i}a},
  {Laguna}, {Mart{\'\i}n}, {Navarro}, {N{\'u}{\~n}ez Peral}, {Royo},
  {S{\'a}nchez}, {Silva-L{\'o}pez}, {Vera}, {Villanueva}, {Fourmond}, {de
  Galarreta}, {Bouzit}, {Hervier}, {Le Clec'h}, {Szwec}, {Chaigneau},
  {Buttice}, {Dominguez-Tagle}, {Philippon}, {Boumier}, {Le Cocguen},
  {Baranjuk}, {Bell}, {Berkefeld}, {Baumgartner}, {Heidecke}, {Maue}, {Nakai},
  {Scheiffelen}, {Sigwarth}, {Soltau}, {Volkmer}, {Blanco Rodr{\'\i}guez},
  {Domingo}, {Ferreres Sabater}, {Gasent Blesa}, {Rodr{\'\i}guez
  Mart{\'\i}nez}, {Osorno Caudel}, {Bosch}, {Casas}, {Carmona}, {Herms},
  {Roma}, {Alonso}, {G{\'o}mez-Sanjuan}, {Piqueras}, {Torralbo}, {Fiethe},
  {Guan}, {Lange}, {Michel}, {Bonet}, {Fahmy}, {M{\"u}ller}, \&
  {Zouganelis}}]{solanki20}
{Solanki}, S.~K., {del Toro Iniesta}, J.~C., {Woch}, J., {et~al.} 2020, \aap,
  642, A11, \dodoi{10.1051/0004-6361/201935325}

\bibitem[{{Sterling} \& {Moore}(2020)}]{sterling20}
{Sterling}, A.~C., \& {Moore}, R.~L. 2020, \apjl, 896, L18,
  \dodoi{10.3847/2041-8213/ab96be}

\bibitem[{{Szente} {et~al.}(2022){Szente}, {Landi}, \& {van der
  Holst}}]{2022ApJ...926...35S}
{Szente}, J., {Landi}, E., \& {van der Holst}, B. 2022, \apj, 926, 35,
  \dodoi{10.3847/1538-4357/ac3918}

\bibitem[{{Telloni} {et~al.}(2022){Telloni}, {Zank}, {Stangalini}, {Downs},
  {Liang}, {Nakanotani}, {Andretta}, {Antonucci}, {Sorriso-Valvo}, {Adhikari},
  {Zhao}, {Marino}, {Susino}, {Grimani}, {Fabi}, {D'Amicis}, {Perrone},
  {Bruno}, {Carbone}, {Mancuso}, {Romoli}, {Deppo}, {Fineschi}, {Heinzel},
  {Moses}, {Naletto}, {Nicolini}, {Spadaro}, {Teriaca}, {Frassati}, {Jerse},
  {Landini}, {Pancrazzi}, {Russano}, {Sasso}, {Biondo}, {Burtovoi}, {Capuano},
  {Casini}, {Casti}, {Chioetto}, {Leo}, {Giarrusso}, {Liberatore}, {Berghmans},
  {Auch{\`e}re}, {Cuadrado}, {Chitta}, {Harra}, {Kraaikamp}, {Long}, {Mandal},
  {Parenti}, {Pelouze}, {Peter}, {Rodriguez}, {Sch{\"u}hle}, {Schwanitz},
  {Smith}, {Verbeeck}, \& {Zhukov}}]{telloni22}
{Telloni}, D., {Zank}, G.~P., {Stangalini}, M., {et~al.} 2022, \apjl, 936, L25,
  \dodoi{10.3847/2041-8213/ac8104}

\bibitem[{{Timothy} {et~al.}(1975){Timothy}, {Krieger}, \&
  {Vaiana}}]{1975SoPh...42..135T}
{Timothy}, A.~F., {Krieger}, A.~S., \& {Vaiana}, G.~S. 1975, \solphys, 42, 135

\bibitem[{{Titov}(2007)}]{2007ApJ...660..863T}
{Titov}, V.~S. 2007, \apj, 660, 863, \dodoi{10.1086/512671}

\bibitem[{{T{\"o}r{\"o}k} {et~al.}(2018){T{\"o}r{\"o}k}, {Downs}, {Linker},
  {Lionello}, {Titov}, {Miki{\'c}}, {Riley}, {Caplan}, \&
  {Wijaya}}]{2018ApJ...856...75T}
{T{\"o}r{\"o}k}, T., {Downs}, C., {Linker}, J.~A., {et~al.} 2018, \apj, 856,
  75, \dodoi{10.3847/1538-4357/aab36d}

\bibitem[{{Upton} \& {Hathaway}(2014)}]{2014ApJ...780....5U}
{Upton}, L., \& {Hathaway}, D.~H. 2014, \apj, 780, 5,
  \dodoi{10.1088/0004-637X/780/1/5}

\bibitem[{{Usmanov} {et~al.}(2011){Usmanov}, {Matthaeus}, {Breech}, \&
  {Goldstein}}]{2011ApJ...727...84U}
{Usmanov}, A.~V., {Matthaeus}, W.~H., {Breech}, B.~A., \& {Goldstein}, M.~L.
  2011, \apj, 727, 84, \dodoi{10.1088/0004-637X/727/2/84}

\bibitem[{{van der Holst} {et~al.}(2014){van der Holst}, {Sokolov}, {Meng},
  {Jin}, {Manchester}, {T{\'o}th}, \& {Gombosi}}]{2014ApJ...782...81V}
{van der Holst}, B., {Sokolov}, I.~V., {Meng}, X., {et~al.} 2014, \apj, 782,
  81, \dodoi{10.1088/0004-637X/782/2/81}

\bibitem[{{Velli}(1993)}]{1993A&A...270..304V}
{Velli}, M. 1993, \aap, 270, 304

\bibitem[{{Verdini} \& {Velli}(2007)}]{2007ApJ...662..669V}
{Verdini}, A., \& {Velli}, M. 2007, \apj, 662, 669, \dodoi{10.1086/510710}

\bibitem[{{Wang} {et~al.}(1996){Wang}, {Hawley}, \&
  {Sheeley}}]{1996Sci...271..464W}
{Wang}, Y.-M., {Hawley}, S.~H., \& {Sheeley}, N.~R. 1996, Science, 271, 464.
\newblock
  \url{http://adsabs.harvard.edu/cgi-bin/nph-bib_query?bibcode=1996Sci...271..464W&db_key=AST}

\bibitem[{{Wang} {et~al.}(1991){Wang}, {Sheeley}, \&
  {Nash}}]{1991ApJ...383..431W}
{Wang}, Y.~M., {Sheeley}, N.~R., J., \& {Nash}, A.~G. 1991, \apj, 383, 431,
  \dodoi{10.1086/170800}

\bibitem[{Webb \& Howard(2012)}]{Webb2012}
Webb, D.~F., \& Howard, T.~A. 2012, Living Reviews in Solar Physics, 9,
  \dodoi{10.12942/lrsp-2012-3}

\bibitem[{{Weinzierl} {et~al.}(2016){Weinzierl}, {Yeates}, {Mackay}, {Henney},
  \& {Arge}}]{weinzierl16}
{Weinzierl}, M., {Yeates}, A.~R., {Mackay}, D.~H., {Henney}, C.~J., \& {Arge},
  C.~N. 2016, \apj, 823, 55, \dodoi{10.3847/0004-637X/823/1/55}

\bibitem[{{Worden} \& {Harvey}(2000)}]{2000SoPh..195..247W}
{Worden}, J., \& {Harvey}, J. 2000, \solphys, 195, 247,
  \dodoi{10.1023/A:1005272502885}

\bibitem[{{Wyper} {et~al.}(2021){Wyper}, {Antiochos}, {DeVore}, {Lynch},
  {Karpen}, \& {Kumar}}]{Wyper2021}
{Wyper}, P.~F., {Antiochos}, S.~K., {DeVore}, C.~R., {et~al.} 2021, \apj, 909,
  54, \dodoi{10.3847/1538-4357/abd9ca}

\bibitem[{{Yeates}(2017)}]{yeates17}
{Yeates}, A.~R. 2017, \apj, 836, 131, \dodoi{10.3847/1538-4357/aa5c84}

\bibitem[{{Yeates} {et~al.}(2018){Yeates}, {Amari}, {Contopoulos}, {Feng},
  {Mackay}, {Miki{\'c}}, {Wiegelmann}, {Hutton}, {Lowder}, {Morgan}, {Petrie},
  {Rachmeler}, {Upton}, {Canou}, {Chopin}, {Downs}, {Druckm{\"u}ller},
  {Linker}, {Seaton}, \& {T{\"o}r{\"o}k}}]{yeates18_ssrv}
{Yeates}, A.~R., {Amari}, T., {Contopoulos}, I., {et~al.} 2018, \ssr, 214, 99

\bibitem[{{Zank} {et~al.}(2012){Zank}, {Jetha}, {Hu}, \&
  {Hunana}}]{2012ApJ...756...21Z}
{Zank}, G.~P., {Jetha}, N., {Hu}, Q., \& {Hunana}, P. 2012, \apj, 756, 21,
  \dodoi{10.1088/0004-637X/756/1/21}

\bibitem[{{Zank} {et~al.}(1996){Zank}, {Matthaeus}, \&
  {Smith}}]{1996JGR...10117093Z}
{Zank}, G.~P., {Matthaeus}, W.~H., \& {Smith}, C.~W. 1996, \jgr, 101, 17093,
  \dodoi{10.1029/96JA01275}

\bibitem[{{Zank} {et~al.}(2020){Zank}, {Nakanotani}, {Zhao}, {Adhikari}, \&
  {Kasper}}]{zank20}
{Zank}, G.~P., {Nakanotani}, M., {Zhao}, L.~L., {Adhikari}, L., \& {Kasper}, J.
  2020, \apj, 903, 1, \dodoi{10.3847/1538-4357/abb828}

\bibitem[{{Zurbuchen} {et~al.}(2002){Zurbuchen}, {Fisk}, {Gloeckler}, \& {von
  Steiger}}]{zurbuchen02_grl}
{Zurbuchen}, T.~H., {Fisk}, L.~A., {Gloeckler}, G., \& {von Steiger}, R. 2002,
  \grl, 29, 66, \dodoi{10.1029/2001GL013946}

\end{thebibliography}

\end{document}